\documentclass[prl,aps,epsfig,superscriptaddress,showpacs,twocolumn,notitlepage]{revtex4-1}
\usepackage{amsmath,amssymb,amsfonts,amstext}
\usepackage{graphicx,color,setspace}
\usepackage[T1]{fontenc}
\usepackage[latin9]{inputenc}
\usepackage[normalem]{ulem}
\usepackage[unicode=true,pdfusetitle,pdfborder={0 0 1}]{hyperref}
\usepackage{upgreek}
\usepackage{bm}
\usepackage{mathrsfs}
\hypersetup{colorlinks,
linkcolor=magenta,          citecolor=blue,        filecolor=blue,      urlcolor=blue           }

\usepackage{times}

\usepackage{epstopdf}

\begin{document}
\title{Artificial Neural Network Based Computation for Out-of-Time-Ordered Correlators}
\author{Yukai Wu}
\affiliation{Center for Quantum Information, IIIS, Tsinghua University, Beijing 100084, P. R. China}
\author{L.-M. Duan}\email{lmduan@tsinghua.edu.cn}
\affiliation{Center for Quantum Information, IIIS, Tsinghua University, Beijing 100084, P. R. China}
\author{Dong-Ling Deng}\email{dldeng@tsinghua.edu.cn}
\affiliation{Center for Quantum Information, IIIS, Tsinghua University, Beijing 100084, P. R. China}
\date{\today}

\begin{abstract}
Out-of-time-ordered correlators  (OTOCs) are of crucial importance for studying a wide variety of fundamental phenomena in quantum physics, ranging from information scrambling to quantum chaos and many-body localization. However, apart from a few special cases, they are notoriously difficult to compute even numerically due to the exponential complexity of generic quantum many-body systems. In this paper, we introduce a machine learning approach to OTOCs based on the restricted-Boltzmann-machine  architecture, which features wide applicability and could work for arbitrary-dimensional systems with massive entanglement. We show, through a concrete example involving a two-dimensional transverse field Ising model,  that our method is capable of computing early-time OTOCs with respect to random pure quantum states or infinite-temperature thermal ensembles. Our results showcase the great potential for machine learning techniques in computing OTOCs, which open up numerous directions for future studies related to similar physical quantities.

%
%
%
\end{abstract}

\maketitle

\begin{figure}[!tbp]
  \centering
  \includegraphics[width=0.8\linewidth]{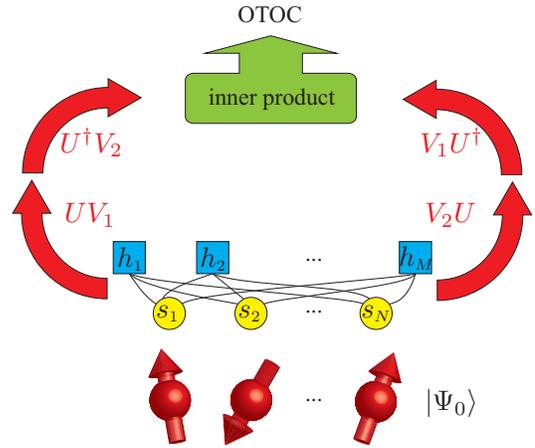}\\
  \caption{A pictorial illustration for the computation of out-of-time-ordered correlators (OTOCs) by using restricted Boltzmann machines (RBMs). Considering an arbitrary state $|\Psi_0\rangle$,  the OTOC as defined in Eq. (\ref{eq:otoc}) can be regarded as the overlap between two states $|\Psi_1\rangle = U^\dag V_2 U V_1|\Psi_0\rangle$ and $|\Psi_2\rangle = V_1 U^\dag V_2 U |\Psi_0\rangle$ (where $U=\exp(-iHt)$ is the time-evolution operator),  which can be efficiently calculated in the RBM represtentaion.}\label{fig:scheme}
\end{figure}

Out-of-time-ordered correlators (OTOCs), first introduced by Larkin and Ovchinnikov in the context of superconductivity \cite{larkin1969quasiclassical}, have attracted tremendous attention across different communities, including quantum information, high-energy physics, and condensed matter physics.
Through analytical and numerical studies, OTOCs of various many-body quantum systems have been computed to characterize their properties in quantum information scrambling \cite{Hosur2016,Mezei2017,PhysRevX.8.021013,PhysRevX.8.021014,PhysRevLett.117.091602,bohrdt2017scrambling,doi:10.1002/andp.201600318,PhysRevB.95.060201,1608.02765,PhysRevB.95.165136,PhysRevA.99.052322}, quantum chaos \cite{PhysRevLett.115.131603,Hosur2016,Maldacena2016,PhysRevB.96.060301} and equilibrium and dynamical quantum phase transitions \cite{PhysRevLett.121.016801,FAN2017707,PhysRevLett.123.140602}. In addition, it has been shown that OTOCs would shed new light on the study of quantum gravity and black holes via AdS/CFT duality \cite{Maldacena1999,Sekino_2008,Shenker2014,Shenker2014_2,Roberts2015,Shenker2015,PhysRevLett.117.111601,doi:10.1002/prop.201700034,1710.03363}. Recently, OTOCs have also been experimentally measured  in systems of trapped ions \cite{garttner2017measuring}, solid-state spins \cite{PhysRevX.7.031011,PhysRevLett.120.070501}, and  $^{87}$Rb Bose-Einstein condensate \cite{PhysRevA.100.013623}, etc. 
Here, we introduce machine learning, an important tool borrowed from computer science \cite{Michalski2013Machine, LeCun2015Deep, Jordan2015Machine} , to the studies  of OTOCs, with focus on numerical computation of  OTOCs by using  restricted Boltzmann machines (RBMs) (see Fig.\ref{fig:scheme} for a schematic illustration).

Apart from
some analytically solvable examples (e.g. \cite{kitaev2015otoc,Polchinski2016,PhysRevD.94.106002,PhysRevLett.122.020603,PhysRevB.97.144304})
, the numerical computation of OTOCs for generic quantum many-body systems is notoriously challenging  due to the exponential scaling of the Hilbert space dimension with the system size. In one-dimensional (1D) systems with short-range interaction, OTOCs can be computed using tensor network methods such as time evolving block decimation (TEBD)
\cite{PhysRevLett.121.016801}
and matrix product operators (MPO) \cite{1802.00801}. However, once long-range interactions are included, these methods may no longer be efficient since the entanglement in the system grows quickly and there is no apparent way to write down local MPOs \cite{Schollwock2011Density}. In higher dimensions, the tensor contraction is a $\#$P-complete problem \cite{garcia2011exact}, rendering most of the traditional tensor-network based methods unfeasible as well.

In this paper, we propose a machine-learning approach for evaluating early-time OTOCs that would bypass these difficulties and  work for arbitrary-dimensional systems with massive entanglement. We mention that within physics, applications of machine-learning techniques have recently been boosted in a number of different contexts \cite{carleo2019machine,Sarma2019Machine,Carleo2016Solving,Deng2017Quantum,Arsenault2015Machine,Zhang2016Triangular,Carrasquilla2017Machine,
van2017Learning,Deng2017Machine,Wang2016Discovering,Broecker2017Machine,
Chng2017Machine,Zhang2017Machine,Wetzel2017Unsupervised,Hu2017Discovering,
Yoshioka2017Learning,Torlai2016Learning,Aoki2016restricted,You2017Machine,
Torlai2018Neural,Pasquato2016Detecting,Hezaveh2017Fast,Rahul2013Application,
Abbott2016Observation,Kalinin2015Big,Schoenholz2016Structural,Liu2017Self,
Huang2017Accelerated,Torlai2017Neural,Gao2017Efficient,Chen2017Equivalence,
Huang2017Neural,Schindler2017Probing,Cai2017Approximating,Broecker2017Quantum,
Nomura2017Restricted,Biamonte2017Quantum,Lu2017Separability,Weinstein2017Learning,Saito2017Solving,
Saito2017Machine,Schmitt2017Quantum,Deng2017MachineBN,Hsu2018Machine}, including material design \cite{Kalinin2015Big}, gravitational
lenses \cite{Hezaveh2017Fast} and wave analysis \cite{Rahul2013Application,Abbott2016Observation}, black hole detection \cite{Pasquato2016Detecting},
 glassy dynamics \cite{Schoenholz2016Structural}, quantum nonlocality detection \cite{Deng2017MachineBN},
 topological
codes \cite{Torlai2017Neural}, quantum machine learning \cite{Biamonte2017Quantum,Sarma2019Machine},
and topological phases and phase transitions \cite{Zhang2016Triangular,Carrasquilla2017Machine,van2017Learning,Wang2016Discovering,
Broecker2017Machine,Chng2017Machine,Zhang2017Machine,Deng2017Machine,
Wetzel2017Unsupervised,Hu2017Discovering},
etc.  Here, we focus on one of the simplest stochastic neural networks
for unsupervised learning\textemdash the restricted Boltzmann machine
 \cite{Hinton2006Reducing,Salakhutdinov2007Restricted,Larochelle2008Classification} and introduce an RBM-based approach to the numerical computation of OTOCs in quantum many-body systems. Through a concrete example of a 2D transverse field Ising model with system sizes as large as ten-by-ten (which is far beyond the capacity of exact diagonalization), we demonstrate that the proposed RBM-based approach is capable of computing the early-time OTOCs with respect to random pure quantum states or infinite-temperature thermal ensembles.  Our method works for generic systems, independent of dimensionality, the amount of entanglement involved,
 or whether the calculation is performed for regions far away from or near the quantum phase transition point.
  Our results showcase the unparalleled
power of machine learning in the studies of OTOCs
for quantum many-body systems, which paves a novel way to study numerous physical
phenomena related to the properties of OTOCs.

\textit{The RBM approach}.---To begin with, let us first briefly introduce the definition of OTOC and the RBM representation of quantum many-body states. An OTOC of a quantum system with Hamiltonian $H$ is defined as \cite{swingle2018unscrambling}
\begin{equation}
\label{eq:otoc}
F(t) = \langle V_2^\dag(t) V_1^\dag V_2(t) V_1 \rangle,
\end{equation}
where $V_1$ and $V_2$ are two quantum operators and $V_2(t) = \exp(iHt)V_2\exp(-iHt)$ is the time-evolved operator in the Heisenberg picture. The expectation value in the above equation can be evaluated with respect to a certain quantum state $|\psi\rangle$, such as the ground state of $H$ or a state that can be easily prepared in the experiment (e.g. Ref.~\cite{PhysRevLett.121.016801,garttner2017measuring}), or with respect to an ensemble of states, such as a thermal ensemble at temperature $T$ (e.g. Ref.~\cite{Hosur2016,Mezei2017}). From the definition in Eq. (\ref{eq:otoc}), the evaluation of the OTOC involves the action of operators on the quantum states, their time evolution, and the overlap between different states. In general, the states and operators of an $N$-qubit system are represented by vectors and matrices of dimension $2^N$, whose storage and manipulation require a formidably huge amount of computational resources when $N$ is large. This is the major challenge in the numerical evaluation of OTOCs and many other quantities of quantum many-body systems. Fortunately, in practice the physical states and operators we are interested in typically have certain structures and only occupy a tiny corner of the entire Hilbert space, hence possibly allowing much more efficient representations.

One of such efficient representations is the RBM representation, which has attracted considerable attention recently \cite{Carleo2016Solving,carleo2019machine, Deng2017Quantum,Deng2017Machine,Deng2017MachineBN}. In an RBM representation, a system of $N$ spins can be represented by a set of network parameters $\{a,\,b,\,W\}$. The (unnormalized) many-body wavefunction is given by
\begin{equation}
\Psi(S) = \sum_{\{h_i\}} \exp\left( \sum_j a_j s_j + \sum_i b_i h_i + \sum_{ij} W_{ij} h_i s_j \right), \label{eq:wavefunction}
\end{equation}
where $S=(s_1,\,\cdots,\,s_N)$ with $s_j=\pm 1$ represents a spin configuration in the $\sigma_z$ basis; $h=(h_1,\,\cdots,\,h_M)$ with $h_i=\pm 1$ describes the state of $M$ hidden spin variables; $a$ and $b$ are $N$- and $M$- dimensional complex vectors, and $W$ an $M\times N$ complex matrix. After tracing out the hidden spin variables explicitly, the wavefunction can also be written as $
\Psi(S) = \exp\left(\sum_j a_j s_j\right)  \prod_i 2\cosh \left( b_i + \sum_j W_{ij} s_j \right)$. We mention that any quantum state can
be approximated to arbitrary accuracy by the above RBM representation,
as long as the number of hidden neurons is large enough \cite{Kolmogorov1963Representation,Le2008Representational,Hornik1991Approximation}.


Now we introduce the general recipe for computing OTOCs by using the RBM representaiton.
The basic procedure is illustrated schematically in Fig.~\ref{fig:scheme}. Suppose the expectation value in Eq.~(\ref{eq:otoc}) is with respect to a quantum state $|\Psi_0\rangle$ represented by an RBM. First, we rewrite the OTOC as the overlap between two states $|\Psi_1\rangle = V_2(t)V_1|\Psi_0\rangle$ and $|\Psi_2\rangle = V_1 V_2(t)|\Psi_0\rangle$. Then, we plug in the expression for $V_2(t)$ and get $|\Psi_1\rangle = U^\dag V_2 U V_1|\Psi_0\rangle$ and $|\Psi_2\rangle = V_1 U^\dag V_2 U |\Psi_0\rangle$ where $U=\exp(-iHt)$ is the time-evolution operator for time $t$. Physically it means that the state $|\Psi_1\rangle$ comes from the initial state $|\Psi_0\rangle$, acted on by an operator $V_1$, time-evolved for a time interval of $t$, further acted on by an operator $V_2$, and finally time-evolved backwards for a time interval of $t$; and similar interpretation can be given for the state $|\Psi_2\rangle$. Therefore, in order to compute the OTOC, it is of crucial importance that we should be able to find efficient RBM representation of the initial state, to describe the action of operators $V_1$ and $V_2$ on an RBM state, to solve the time-evolution of an RBM state, and finally to compute the overlap between two RBM states.

%

In the definition of OTOC, $V_1$ and $V_2$ can be arbitrary operators. Here, we focus on local Pauli operators, which is a natural choice for spin systems and has been widely used in the studies of quantum chaos and   information scrambling \cite{Mezei2017,doi:10.1002/andp.201600318,PhysRevB.95.060201,PhysRevA.99.052322}. Let us consider how the operator $\sigma_\alpha^k$ ($\alpha=x,\,y,\,z$ and $k=1,\,2,\,\cdots,\,N$) acts on an RBM state described by the parameters $\{a,\,b,\,W\}$. The effect of $\sigma_x^k$ is to flip the $k$-th spin. In other words, we want to replace $s_k$ in Eq.~(\ref{eq:wavefunction}) by $-s_k$ while keeping the value of the wave function unchanged. This can be achieved by updating the RBM parameters $a_k\to -a_k$ and $W_{ik}\to -W_{ik}$ ($i=1,\,\cdots,\,M$). Therefore, to get the new state after the application of $\sigma_x^k$, we only need to update $(M+1)$ parameters instead of dealing with the $2^N$-dimensional state vectors. Next we consider $\sigma_z^k$. Its action on an RBM state  will not change the spin configuration in the $\sigma_z$ basis,  but introduce a relative phase of $\pi$ between the $s_k=\pm 1$ states. Therefore,  we can efficiently describe the resulted state after applying $\sigma_z^k$ by updating $a_k\to a_k+i\pi/2$. Note that in this way we get a global phase factor $i$ in addition to the desired effect of applying $\sigma_z^k$ operator. This additional phase factor must be carefully treated when computing the overlap between general states.  However,  in our calculation of OTOCs, the same operator appears in both $|\Psi_1\rangle$ and $|\Psi_2\rangle$ and therefore the phase factors cancel with each other in the inner product. Based on this, we can further implement the $\sigma_y^k$ operator by consecutive actions of $\sigma_x^k$ and $\sigma_z^k$, without worrying about the global phase factor.

The time evolution of an RBM state can be performed in a similar way as training the ground state \cite{Carleo2016Solving}. At each step, we try to maximize the fidelity between a new RBM state $|\Psi(t+\delta t)\rangle$ and the time-evolved state $(I-i H \delta t)|\Psi(t)\rangle$. For simplicity in notation here we assume that the states are normalized. Actually, according to Ref.~\cite{Carleo2016Solving}, this optimization can be realized by simply using an imaginary learning rate in the algorithm for the ground state. However, in this way we may get an additional phase factor $|\Psi(t+\delta t)\rangle=e^{i\delta\phi}(I-i H \delta t)|\Psi(t)\rangle$. Such a global phase is irrelevant in evaluating expectation values since the same phase and its complex conjugate will cancel each other; but for OTOC, we need to compute the overlap between two quantum states, hence we must keep track of all the phase changes during the evolution. Specifically, we compute the overlap between the states before and after the evolution at each step and get $\langle\Psi(t)|\Psi(t+\delta t)\rangle=e^{i\delta\phi}(1-i \bar{E}(t) \delta t)$ where $\bar{E}(t)$ is the average energy at time $t$. In this way we can get the phase shift at each step and then remove them from the final OTOC calculation.

Finally, we consider the overlap between two RBM states $\langle\Psi_2|\Psi_1\rangle / \sqrt{\langle\Psi_1|\Psi_1\rangle\langle\Psi_2|\Psi_2\rangle}$, which can be obtained by Monte Carlo sampling of the spin configurations. Specifically, we can sample the spin configuration $S$ for $|\Psi_1\rangle$ with relative probability $|\langle S|\Psi_1\rangle|^2$, or normalized probability $\langle\Psi_1|S\rangle\langle S|\Psi_1\rangle/\langle\Psi_1|\Psi_1\rangle$. For each sampled spin configuration, we compute $\langle S|\Psi_2\rangle/\langle S|\Psi_1\rangle$. By averaging over all the spin configurations, we get $v_1=\sum_S\langle\Psi_1|S\rangle\langle S|\Psi_2\rangle/\langle\Psi_1|\Psi_1\rangle = \langle\Psi_1|\Psi_2\rangle/\langle\Psi_1|\Psi_1\rangle$. Similarly we can sample for $|\Psi_2\rangle$ and compute $\langle S|\Psi_1\rangle/\langle S|\Psi_2\rangle$. The average value we get is $v_2= \langle\Psi_2|\Psi_1\rangle/\langle\Psi_2|\Psi_2\rangle$. Combining the two results together we get $\langle \Psi_2|\Psi_1\rangle/\sqrt{\langle\Psi_1|\Psi_1\rangle\langle\Psi_2|\Psi_2\rangle} = \sqrt{v_1^* v_2}$.

We stress the difference between our RBM and the conventional TEBD or MPO (or more general tensor-network based) approaches to computing OTOCs. Generally speaking, the TEBD approach relies vitally on the efficient matrix-product-state representation of a quantum many-body state, hence is limited to 1D systems with  short-range interactions and small entanglement \cite{Schollwock2011Density}.  The MPO approach exploits the fact that the operators (in the Heisenberg picture) expand at most ballistically for local Hamiltonians with speed bounded by the Lieb-Robinson speed \cite{1802.00801}, thus is applicable to a much wider space time region than the TEBD approach. Yet, for systems in higher dimensions (larger than one) or with long-range interactions, the MPO approach suffers still since tensor contraction is inefficient in higher dimensions and there is no apparent way to write down local MPOs for systems with long-range interactions. In stark contrast, our RBM approach escapes these limitations owing to the particular neural network structures. It works for higher dimensions and long-range interactions. In addition, since entanglement is not a limiting factor for the efficiency of the RBM representation \cite{Deng2017Quantum}, we expect that it can be used to computing OTOCs for quantum states with massive (e.g., volume-law) entanglement as well. To show more precisely how this RBM approach works, we give a concrete example involving computing OTOCs for a 2D transverse field Ising model, which is beyond the capacity of the TEBD or MPO methods for large system sizes.

\begin{figure}[!tbp]
\centering
  \includegraphics[width=0.96\linewidth]{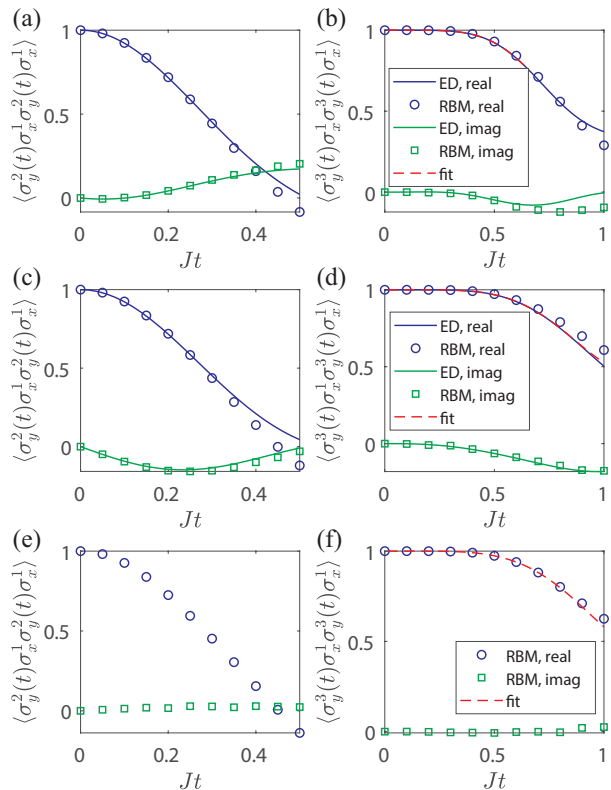}\\
  \caption{Comparison between restricted Boltzmann machine (RBM) and exact diagonalization (ED) results for $N=3\times 4$ (first row), $N=4\times 5$ (second row) and $N=10\times 10$ (third row) spins in a transverse field Ising model at $h/J=1$. Here we choose $V_1=\sigma_x$ and $V_2=\sigma_y$ on the nearest neighbor (first column) and second nearest neighbor (second column) sites. As we can see, both the real and imaginary parts of the RBM result agree well with ED in the early time, and starts to deviate when the OTOC is significantly away from the initial value of one. For the second nearest neighbor case, we fit the early time behavior according to Eq.~(\ref{eq:fit}) and consistently obtain $\lambda\approx 1.9$, $v_f\approx 2.0$ and $p\approx 0.44$ for different system sizes. }\label{fig:compare_h1}
\end{figure}

\textit{A 2D example}.---We consider a $2D$ transverse field Ising  model on an $N=L_1\times L_2$ square lattice with periodic boundary conditions. The Hamiltonian of the system is given by:
\begin{equation}
H = -h\sum_i \sigma_x^i - J\sum_{\langle i,\,j\rangle} \sigma_z^i \sigma_z^j,
\end{equation}
where $\langle i,\,j\rangle$ denotes all the nearest neighbor spin pairs. This Hamiltonian is rotated by $90^\circ$ from the commonly used convention \cite{PhysRevA.82.062313}, with $\sigma_x$ and $\sigma_z$ exchanged for convenience. This model is one of the simplest toy models for studying quantum phase transitions, despite the fact that it has the same complexity as a 3D classical Ising model whose exact solution still remains a major open question in statistical physics \cite{chakrabarti2008quantum}.  At zero temperature, a quantum phase transition occurs at $h/J\approx 3$ according to previous studies \cite{PhysRevA.82.062313}.

Here, we compute the OTOCs for the above 2D Ising model by using our  introduced RBM approach.  First we consider random initial RBM states with $\{a,\,b,\,W\}$ following a normal distribution $N(0,\,\sigma^2)$ for their real and imaginary parts. For small system sizes we use $\sigma=0.1$ while later for a larger system we reduce it to $\sigma=0.02$ so as to get better performance for the training of RBM \cite{Hinton2012}.
In Fig.~\ref{fig:compare_h1}(a-d), we show the OTOC results for two system sizes $N=3\times 4$ and $N=4\times 5$ at $h/J=1$, which is away from the phase transition point $h/J\approx 3$. From this figure, it is clear that our RBM results match excellently with these from exact diagonalization for small $Jt$ and deviations become noticeable only after
the OTOCs are significantly away from their initial value. This validates the effectiveness of the RBM approach in computing early-time OTOCs. More strikingly, since the complexity of the RBM approach only scales cubically with increasing $N$, we can use it to compute OTOCs for much larger systems. In Fig. ~\ref{fig:compare_h1}(e, f), we show
part of our OTOC
results for a system as large as ten by ten, which is far beyond the capacity of exact diagonalization.
%

%
%

In Ref. \cite{1802.00801}, Xu and Swingle have conjectured a universal form for the early-time dynamics of OTOCs:
\begin{eqnarray}
\label{eq:fit}
\text{Re}[F(t)]\sim 1- \frac{1}{2}\text{exp}\left(-\lambda\frac{(d-v_ft)^{1+p}}{t^p} \right),
\end{eqnarray}
where $v_f$ is the speed of the  wavefront and $d$ is the distance between $V_1$ and $V_2$; the index $p$ characterizes the spreading of the wavefront, with $p=0$ corresponding to  pure exponential decay of $\text{Re}[F(t)]$ (for holographic models, coupling Sachdev-Ye-Kitaev clusters, etc.), $p=1/2$ for non-interacting particle models, and $p=1$ for the local random circuit models. Here, we test this conjecture with our RBM results. In Fig. \ref{fig:compare_h1}, we fit the early time ($\text{Re}[F(t)]>0.85$) results for all the three system sizes at the distance $d=2$ and consistently get $\lambda\approx 1.9$, $v_f\approx 2.0$ and $p\approx 0.44$. This corresponds to a sub-diffusively spreading wavefront \cite{1802.00801}.


%

\begin{figure}[!tbp]
  \centering
    \includegraphics[width=0.96\linewidth]{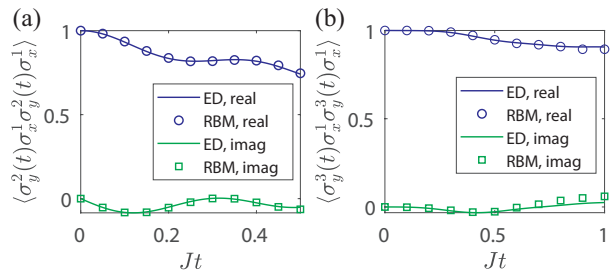}
  \caption{Comparison between restricted Boltzmann machine (RBM) and exact diagonalization (ED) results for $N=4\times 5$ spins at $h/J=3.05$ near the phase transition point of the 2D transverse field Ising model. Again we choose $V_1=\sigma_x$ and $V_2=\sigma_y$ on the (a) nearest neighbor and (b) second nearest neighbor sites. The initial state $|\Psi_0\rangle$ is an RBM state trained to the ground state of $H$.}\label{fig:compare_h3}
\end{figure}

When using RBM-based reinforcement learning to compute the ground state of a Hamiltonian, an observation is that the relative error is usually larger near the critical point due to the divergence of correlation length at the phase transition point \cite{Carleo2016Solving}.
Similar results have also been observed in our calculation of OTOCs. If we take $|\Psi_0\rangle$ to be a random initial state as in Fig.  \ref{fig:compare_h1}, the relative error of OTOCs computed near the phase transition point is larger than that computed deep in the ferromagnetic/paramagnetic phases.
However, if we take $|\Psi_0\rangle$ to be the ground state of the Hamiltonian (e.g. Ref.~\cite{PhysRevLett.121.016801}), we can still observe excellent agreement between RBM and ED methods even at the transition point, as shown in Fig.~\ref{fig:compare_h3} for $N=4\times 5$ spins at $h/J=3.05$.
In addition, we mention that the accuracy of the OTOCs computed via our RBM method can be systematically improved by increasing the number of hidden neurons or iterations in the training process \cite{Carleo2016Solving}.

\begin{figure}[!tbp]
  \centering
  \includegraphics[width=0.96\linewidth]{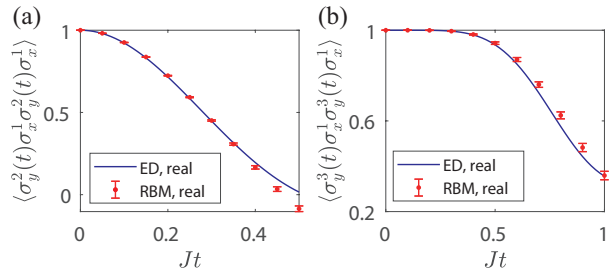}
  \caption{Comparison between restricted Boltzmann machine (RBM) and exact diagonalization (ED) results for $N=3\times 4$ spins at $h/J=1$ and inverse temperature $\beta=0$. Again we consider $V_1=\sigma_x$ and $V_2=\sigma_y$ on the (a) nearest neighbor and (b) second nearest neighbor sites. The RBM results are averaged over $s=10$ randomly generated spin configurations in the $\sigma_x$ basis. The error bars are estimated from the standard deviation of the average values.}\label{fig:thermal}
\end{figure}

In all the above calculations, we compute  OTOCs for pure RBM states. In many theoretical works, the OTOC is evaluated with respect to a thermal distribution at inverse temperature $\beta$. The $\beta=0$ limit corresponds to a uniformly random distribution over all possible spin configurations in our spin model. Since the OTOC is the inner product of two states and thus its absolute value bounded by one, we only need to generate $s$ random spin configurations to upper-bound the accuracy of the average OTOC to $1/\sqrt{s}$. Actually in many cases we are only interested in the real part of OTOC, which is related to the squared out-of-time-ordered commutators \cite{swingle2018unscrambling}. From the previous results we see that it always falls from one and the early-time behavior seems not sensitive to the random choice of initial states; thus we expect the convergence to be much faster. In Fig.~\ref{fig:thermal} we show the infinite temperature OTOC for $N=3\times 4$ and $h/J=1$ by averaging over $s=10$ random spin configurations in the $\sigma_x$ basis, which we generate by training the ground state of Hamiltonian $H=\sum_i \pm \sigma_x^i$. The error bars are estimated from the standard deviation of the average values. As we can see, the small number of random realizations $s=10$ already leads to good convergence; and the discrepancy between the RBM method and the exact results is mainly caused by the representability of RBM states at large time, similar to the previous examples. Note that in practice the performance of the RBM method is weakened if the initial state is exactly a product state. Thus we choose to train the initial states close to the desired states with small randomness, rather than write down an exact solution.

\textit{Discussion and conclusion}.---Although we only focus our discussion on RBMs in this paper,   one may also use other type of neural networks (e.g., deep Boltzmann machine \cite{Gao2017Efficient} or feedforward neural networks \cite{Cai2017Approximating}, etc.) with different learning algorithms to compute OTOCs for different quantum many-body  systems. In particular, it has been proved that deep Boltzmann machine can efficiently represent most physical states, including the ground states of many-body Hamiltonians and states generated by quantum dynamics \cite{Gao2017Efficient}. Therefore, it would be interesting and important to develop a method based on deep Boltzmann machine to compute OTOCs. A complete study on computing OTOCs with different neural networks would not only bring new powerful tools for solving intricate problems in the quantum many-body physics, but also provide valuable insight in understanding the internal  structures of the networks themselves. Moreover, with these new machine learning tools, it would also be interesting and crucial to study certain new physics related to OTOCs, such as information scrambling and dynamical quantum phase transitions in higher dimensions.
We leave these interesting topics for future investigation.

To summarize, in this work we describe a general method of computing OTOC in spin systems using RBM ansatz and then present applications in a 2D transverse field Ising model where numerical calculation was challenging with the existing methods. From our numerical examples, it can be seen that the RBM method is suitable for the early-time properties of OTOC such as the Lyapunov exponent \cite{Shenker2014} and butterfly velocity \cite{Roberts2015,Mezei2017}. The RBM method is not subjected to the limitation of entanglement and geometry, like the conventional method based on local tensor networks, and there is no clear sign problem like the quantum Monte Carlo method. Therefore the RBM method may demonstrate advantages in many models where the other methods are not applicable. On the other hand, what is the limiting factor in the performance of the RBM method is still not clear and can be the topic of future studies.

We thank Jinwu Ye and Shenglong Xu for helpful discussions. This work was supported by the Frontier Science Center for Quantum Information of the Ministry of Education of China, Tsinghua University Initiative Scientific Research Program, and the National key Research and Development Program of China (2016YFA0301902). Y.-K. W. acknowledges support from Shuimu Tsinghua Scholar Program and International Postdoctoral Exchange Fellowship Program (Talent-Introduction Program).


\begin{thebibliography}{96}%
\makeatletter
\providecommand \@ifxundefined [1]{%
 \@ifx{#1\undefined}
}%
\providecommand \@ifnum [1]{%
 \ifnum #1\expandafter \@firstoftwo
 \else \expandafter \@secondoftwo
 \fi
}%
\providecommand \@ifx [1]{%
 \ifx #1\expandafter \@firstoftwo
 \else \expandafter \@secondoftwo
 \fi
}%
\providecommand \natexlab [1]{#1}%
\providecommand \enquote  [1]{``#1''}%
\providecommand \bibnamefont  [1]{#1}%
\providecommand \bibfnamefont [1]{#1}%
\providecommand \citenamefont [1]{#1}%
\providecommand \href@noop [0]{\@secondoftwo}%
\providecommand \href [0]{\begingroup \@sanitize@url \@href}%
\providecommand \@href[1]{\@@startlink{#1}\@@href}%
\providecommand \@@href[1]{\endgroup#1\@@endlink}%
\providecommand \@sanitize@url [0]{\catcode `\\12\catcode `\$12\catcode
  `\&12\catcode `\#12\catcode `\^12\catcode `\_12\catcode `\%12\relax}%
\providecommand \@@startlink[1]{}%
\providecommand \@@endlink[0]{}%
\providecommand \url  [0]{\begingroup\@sanitize@url \@url }%
\providecommand \@url [1]{\endgroup\@href {#1}{\urlprefix }}%
\providecommand \urlprefix  [0]{URL }%
\providecommand \Eprint [0]{\href }%
\providecommand \doibase [0]{http://dx.doi.org/}%
\providecommand \selectlanguage [0]{\@gobble}%
\providecommand \bibinfo  [0]{\@secondoftwo}%
\providecommand \bibfield  [0]{\@secondoftwo}%
\providecommand \translation [1]{[#1]}%
\providecommand \BibitemOpen [0]{}%
\providecommand \bibitemStop [0]{}%
\providecommand \bibitemNoStop [0]{.\EOS\space}%
\providecommand \EOS [0]{\spacefactor3000\relax}%
\providecommand \BibitemShut  [1]{\csname bibitem#1\endcsname}%
\let\auto@bib@innerbib\@empty
\bibitem [{\citenamefont {Larkin}\ and\ \citenamefont
  {Ovchinnikov}(1969)}]{larkin1969quasiclassical}%
  \BibitemOpen
  \bibfield  {author} {\bibinfo {author} {\bibfnamefont {A.}~\bibnamefont
  {Larkin}}\ and\ \bibinfo {author} {\bibfnamefont {Y.~N.}\ \bibnamefont
  {Ovchinnikov}},\ }\bibfield  {title} {\enquote {\bibinfo {title}
  {Quasiclassical method in the theory of superconductivity},}\ }\href@noop {}
  {\bibfield  {journal} {\bibinfo  {journal} {Sov Phys JETP}\ }\textbf
  {\bibinfo {volume} {28}},\ \bibinfo {pages} {1200} (\bibinfo {year}
  {1969})}\BibitemShut {NoStop}%
\bibitem [{\citenamefont {Hosur}\ \emph {et~al.}(2016)\citenamefont {Hosur},
  \citenamefont {Qi}, \citenamefont {Roberts},\ and\ \citenamefont
  {Yoshida}}]{Hosur2016}%
  \BibitemOpen
  \bibfield  {author} {\bibinfo {author} {\bibfnamefont {P.}~\bibnamefont
  {Hosur}}, \bibinfo {author} {\bibfnamefont {X.-L.}\ \bibnamefont {Qi}},
  \bibinfo {author} {\bibfnamefont {D.~A.}\ \bibnamefont {Roberts}}, \ and\
  \bibinfo {author} {\bibfnamefont {B.}~\bibnamefont {Yoshida}},\ }\bibfield
  {title} {\enquote {\bibinfo {title} {Chaos in quantum channels},}\ }\href
  {\doibase 10.1007/JHEP02(2016)004} {\bibfield  {journal} {\bibinfo  {journal}
  {Journal of High Energy Physics}\ }\textbf {\bibinfo {volume} {2016}},\
  \bibinfo {pages} {4} (\bibinfo {year} {2016})}\BibitemShut {NoStop}%
\bibitem [{\citenamefont {Mezei}\ and\ \citenamefont
  {Stanford}(2017)}]{Mezei2017}%
  \BibitemOpen
  \bibfield  {author} {\bibinfo {author} {\bibfnamefont {M.}~\bibnamefont
  {Mezei}}\ and\ \bibinfo {author} {\bibfnamefont {D.}~\bibnamefont
  {Stanford}},\ }\bibfield  {title} {\enquote {\bibinfo {title} {On
  entanglement spreading in chaotic systems},}\ }\href {\doibase
  10.1007/JHEP05(2017)065} {\bibfield  {journal} {\bibinfo  {journal} {Journal
  of High Energy Physics}\ }\textbf {\bibinfo {volume} {2017}},\ \bibinfo
  {pages} {65} (\bibinfo {year} {2017})}\BibitemShut {NoStop}%
\bibitem [{\citenamefont {von Keyserlingk}\ \emph {et~al.}(2018)\citenamefont
  {von Keyserlingk}, \citenamefont {Rakovszky}, \citenamefont {Pollmann},\ and\
  \citenamefont {Sondhi}}]{PhysRevX.8.021013}%
  \BibitemOpen
  \bibfield  {author} {\bibinfo {author} {\bibfnamefont {C.~W.}\ \bibnamefont
  {von Keyserlingk}}, \bibinfo {author} {\bibfnamefont {T.}~\bibnamefont
  {Rakovszky}}, \bibinfo {author} {\bibfnamefont {F.}~\bibnamefont {Pollmann}},
  \ and\ \bibinfo {author} {\bibfnamefont {S.~L.}\ \bibnamefont {Sondhi}},\
  }\bibfield  {title} {\enquote {\bibinfo {title} {Operator hydrodynamics,
  otocs, and entanglement growth in systems without conservation laws},}\
  }\href {\doibase 10.1103/PhysRevX.8.021013} {\bibfield  {journal} {\bibinfo
  {journal} {Phys. Rev. X}\ }\textbf {\bibinfo {volume} {8}},\ \bibinfo {pages}
  {021013} (\bibinfo {year} {2018})}\BibitemShut {NoStop}%
\bibitem [{\citenamefont {Nahum}\ \emph {et~al.}(2018)\citenamefont {Nahum},
  \citenamefont {Vijay},\ and\ \citenamefont {Haah}}]{PhysRevX.8.021014}%
  \BibitemOpen
  \bibfield  {author} {\bibinfo {author} {\bibfnamefont {A.}~\bibnamefont
  {Nahum}}, \bibinfo {author} {\bibfnamefont {S.}~\bibnamefont {Vijay}}, \ and\
  \bibinfo {author} {\bibfnamefont {J.}~\bibnamefont {Haah}},\ }\bibfield
  {title} {\enquote {\bibinfo {title} {Operator spreading in random unitary
  circuits},}\ }\href {\doibase 10.1103/PhysRevX.8.021014} {\bibfield
  {journal} {\bibinfo  {journal} {Phys. Rev. X}\ }\textbf {\bibinfo {volume}
  {8}},\ \bibinfo {pages} {021014} (\bibinfo {year} {2018})}\BibitemShut
  {NoStop}%
\bibitem [{\citenamefont {Roberts}\ and\ \citenamefont
  {Swingle}(2016)}]{PhysRevLett.117.091602}%
  \BibitemOpen
  \bibfield  {author} {\bibinfo {author} {\bibfnamefont {D.~A.}\ \bibnamefont
  {Roberts}}\ and\ \bibinfo {author} {\bibfnamefont {B.}~\bibnamefont
  {Swingle}},\ }\bibfield  {title} {\enquote {\bibinfo {title} {Lieb-robinson
  bound and the butterfly effect in quantum field theories},}\ }\href {\doibase
  10.1103/PhysRevLett.117.091602} {\bibfield  {journal} {\bibinfo  {journal}
  {Phys. Rev. Lett.}\ }\textbf {\bibinfo {volume} {117}},\ \bibinfo {pages}
  {091602} (\bibinfo {year} {2016})}\BibitemShut {NoStop}%
\bibitem [{\citenamefont {Bohrdt}\ \emph {et~al.}(2017)\citenamefont {Bohrdt},
  \citenamefont {Mendl}, \citenamefont {Endres},\ and\ \citenamefont
  {Knap}}]{bohrdt2017scrambling}%
  \BibitemOpen
  \bibfield  {author} {\bibinfo {author} {\bibfnamefont {A.}~\bibnamefont
  {Bohrdt}}, \bibinfo {author} {\bibfnamefont {C.~B.}\ \bibnamefont {Mendl}},
  \bibinfo {author} {\bibfnamefont {M.}~\bibnamefont {Endres}}, \ and\ \bibinfo
  {author} {\bibfnamefont {M.}~\bibnamefont {Knap}},\ }\bibfield  {title}
  {\enquote {\bibinfo {title} {Scrambling and thermalization in a diffusive
  quantum many-body system},}\ }\href {\doibase 10.1088/1367-2630/aa719b}
  {\bibfield  {journal} {\bibinfo  {journal} {New Journal of Physics}\ }\textbf
  {\bibinfo {volume} {19}},\ \bibinfo {pages} {063001} (\bibinfo {year}
  {2017})}\BibitemShut {NoStop}%
\bibitem [{\citenamefont {Huang}\ \emph {et~al.}(2017)\citenamefont {Huang},
  \citenamefont {Zhang},\ and\ \citenamefont
  {Chen}}]{doi:10.1002/andp.201600318}%
  \BibitemOpen
  \bibfield  {author} {\bibinfo {author} {\bibfnamefont {Y.}~\bibnamefont
  {Huang}}, \bibinfo {author} {\bibfnamefont {Y.-L.}\ \bibnamefont {Zhang}}, \
  and\ \bibinfo {author} {\bibfnamefont {X.}~\bibnamefont {Chen}},\ }\bibfield
  {title} {\enquote {\bibinfo {title} {Out-of-time-ordered correlators in
  many-body localized systems},}\ }\href {\doibase 10.1002/andp.201600318}
  {\bibfield  {journal} {\bibinfo  {journal} {Annalen der Physik}\ }\textbf
  {\bibinfo {volume} {529}},\ \bibinfo {pages} {1600318} (\bibinfo {year}
  {2017})}\BibitemShut {NoStop}%
\bibitem [{\citenamefont {Swingle}\ and\ \citenamefont
  {Chowdhury}(2017)}]{PhysRevB.95.060201}%
  \BibitemOpen
  \bibfield  {author} {\bibinfo {author} {\bibfnamefont {B.}~\bibnamefont
  {Swingle}}\ and\ \bibinfo {author} {\bibfnamefont {D.}~\bibnamefont
  {Chowdhury}},\ }\bibfield  {title} {\enquote {\bibinfo {title} {Slow
  scrambling in disordered quantum systems},}\ }\href {\doibase
  10.1103/PhysRevB.95.060201} {\bibfield  {journal} {\bibinfo  {journal} {Phys.
  Rev. B}\ }\textbf {\bibinfo {volume} {95}},\ \bibinfo {pages} {060201}
  (\bibinfo {year} {2017})}\BibitemShut {NoStop}%
\bibitem [{\citenamefont {Chen}(2016)}]{1608.02765}%
  \BibitemOpen
  \bibfield  {author} {\bibinfo {author} {\bibfnamefont {Y.}~\bibnamefont
  {Chen}},\ }\href@noop {} {\enquote {\bibinfo {title} {Universal logarithmic
  scrambling in many body localization},}\ } (\bibinfo {year} {2016}),\ \Eprint
  {http://arxiv.org/abs/arXiv:1608.02765} {arXiv:1608.02765} \BibitemShut
  {NoStop}%
\bibitem [{\citenamefont {Slagle}\ \emph {et~al.}(2017)\citenamefont {Slagle},
  \citenamefont {Bi}, \citenamefont {You},\ and\ \citenamefont
  {Xu}}]{PhysRevB.95.165136}%
  \BibitemOpen
  \bibfield  {author} {\bibinfo {author} {\bibfnamefont {K.}~\bibnamefont
  {Slagle}}, \bibinfo {author} {\bibfnamefont {Z.}~\bibnamefont {Bi}}, \bibinfo
  {author} {\bibfnamefont {Y.-Z.}\ \bibnamefont {You}}, \ and\ \bibinfo
  {author} {\bibfnamefont {C.}~\bibnamefont {Xu}},\ }\bibfield  {title}
  {\enquote {\bibinfo {title} {Out-of-time-order correlation in marginal
  many-body localized systems},}\ }\href {\doibase 10.1103/PhysRevB.95.165136}
  {\bibfield  {journal} {\bibinfo  {journal} {Phys. Rev. B}\ }\textbf {\bibinfo
  {volume} {95}},\ \bibinfo {pages} {165136} (\bibinfo {year}
  {2017})}\BibitemShut {NoStop}%
\bibitem [{\citenamefont {Da\ifmmode~\breve{g}\else \u{g}\fi{}}\ and\
  \citenamefont {Duan}(2019)}]{PhysRevA.99.052322}%
  \BibitemOpen
  \bibfield  {author} {\bibinfo {author} {\bibfnamefont {C.~B.}\ \bibnamefont
  {Da\ifmmode~\breve{g}\else \u{g}\fi{}}}\ and\ \bibinfo {author}
  {\bibfnamefont {L.-M.}\ \bibnamefont {Duan}},\ }\bibfield  {title} {\enquote
  {\bibinfo {title} {Detection of out-of-time-order correlators and information
  scrambling in cold atoms: Ladder-$\mathit{XX}$ model},}\ }\href {\doibase
  10.1103/PhysRevA.99.052322} {\bibfield  {journal} {\bibinfo  {journal} {Phys.
  Rev. A}\ }\textbf {\bibinfo {volume} {99}},\ \bibinfo {pages} {052322}
  (\bibinfo {year} {2019})}\BibitemShut {NoStop}%
\bibitem [{\citenamefont {Roberts}\ and\ \citenamefont
  {Stanford}(2015)}]{PhysRevLett.115.131603}%
  \BibitemOpen
  \bibfield  {author} {\bibinfo {author} {\bibfnamefont {D.~A.}\ \bibnamefont
  {Roberts}}\ and\ \bibinfo {author} {\bibfnamefont {D.}~\bibnamefont
  {Stanford}},\ }\bibfield  {title} {\enquote {\bibinfo {title} {Diagnosing
  chaos using four-point functions in two-dimensional conformal field
  theory},}\ }\href {\doibase 10.1103/PhysRevLett.115.131603} {\bibfield
  {journal} {\bibinfo  {journal} {Phys. Rev. Lett.}\ }\textbf {\bibinfo
  {volume} {115}},\ \bibinfo {pages} {131603} (\bibinfo {year}
  {2015})}\BibitemShut {NoStop}%
\bibitem [{\citenamefont {Maldacena}\ \emph {et~al.}(2016)\citenamefont
  {Maldacena}, \citenamefont {Shenker},\ and\ \citenamefont
  {Stanford}}]{Maldacena2016}%
  \BibitemOpen
  \bibfield  {author} {\bibinfo {author} {\bibfnamefont {J.}~\bibnamefont
  {Maldacena}}, \bibinfo {author} {\bibfnamefont {S.~H.}\ \bibnamefont
  {Shenker}}, \ and\ \bibinfo {author} {\bibfnamefont {D.}~\bibnamefont
  {Stanford}},\ }\bibfield  {title} {\enquote {\bibinfo {title} {A bound on
  chaos},}\ }\href {\doibase 10.1007/JHEP08(2016)106} {\bibfield  {journal}
  {\bibinfo  {journal} {Journal of High Energy Physics}\ }\textbf {\bibinfo
  {volume} {2016}},\ \bibinfo {pages} {106} (\bibinfo {year}
  {2016})}\BibitemShut {NoStop}%
\bibitem [{\citenamefont {Kukuljan}\ \emph {et~al.}(2017)\citenamefont
  {Kukuljan}, \citenamefont {Grozdanov},\ and\ \citenamefont
  {Prosen}}]{PhysRevB.96.060301}%
  \BibitemOpen
  \bibfield  {author} {\bibinfo {author} {\bibfnamefont {I.}~\bibnamefont
  {Kukuljan}}, \bibinfo {author} {\bibfnamefont {S.~c.~v.}\ \bibnamefont
  {Grozdanov}}, \ and\ \bibinfo {author} {\bibfnamefont {T.~c.~v.}\
  \bibnamefont {Prosen}},\ }\bibfield  {title} {\enquote {\bibinfo {title}
  {Weak quantum chaos},}\ }\href {\doibase 10.1103/PhysRevB.96.060301}
  {\bibfield  {journal} {\bibinfo  {journal} {Phys. Rev. B}\ }\textbf {\bibinfo
  {volume} {96}},\ \bibinfo {pages} {060301} (\bibinfo {year}
  {2017})}\BibitemShut {NoStop}%
\bibitem [{\citenamefont {Heyl}\ \emph {et~al.}(2018)\citenamefont {Heyl},
  \citenamefont {Pollmann},\ and\ \citenamefont
  {D\'ora}}]{PhysRevLett.121.016801}%
  \BibitemOpen
  \bibfield  {author} {\bibinfo {author} {\bibfnamefont {M.}~\bibnamefont
  {Heyl}}, \bibinfo {author} {\bibfnamefont {F.}~\bibnamefont {Pollmann}}, \
  and\ \bibinfo {author} {\bibfnamefont {B.}~\bibnamefont {D\'ora}},\
  }\bibfield  {title} {\enquote {\bibinfo {title} {Detecting equilibrium and
  dynamical quantum phase transitions in ising chains via out-of-time-ordered
  correlators},}\ }\href {\doibase 10.1103/PhysRevLett.121.016801} {\bibfield
  {journal} {\bibinfo  {journal} {Phys. Rev. Lett.}\ }\textbf {\bibinfo
  {volume} {121}},\ \bibinfo {pages} {016801} (\bibinfo {year}
  {2018})}\BibitemShut {NoStop}%
\bibitem [{\citenamefont {Fan}\ \emph {et~al.}(2017)\citenamefont {Fan},
  \citenamefont {Zhang}, \citenamefont {Shen},\ and\ \citenamefont
  {Zhai}}]{FAN2017707}%
  \BibitemOpen
  \bibfield  {author} {\bibinfo {author} {\bibfnamefont {R.}~\bibnamefont
  {Fan}}, \bibinfo {author} {\bibfnamefont {P.}~\bibnamefont {Zhang}}, \bibinfo
  {author} {\bibfnamefont {H.}~\bibnamefont {Shen}}, \ and\ \bibinfo {author}
  {\bibfnamefont {H.}~\bibnamefont {Zhai}},\ }\bibfield  {title} {\enquote
  {\bibinfo {title} {Out-of-time-order correlation for many-body
  localization},}\ }\href {\doibase https://doi.org/10.1016/j.scib.2017.04.011}
  {\bibfield  {journal} {\bibinfo  {journal} {Science Bulletin}\ }\textbf
  {\bibinfo {volume} {62}},\ \bibinfo {pages} {707 } (\bibinfo {year}
  {2017})}\BibitemShut {NoStop}%
\bibitem [{\citenamefont {Da\ifmmode~\breve{g}\else \u{g}\fi{}}\ \emph
  {et~al.}(2019)\citenamefont {Da\ifmmode~\breve{g}\else \u{g}\fi{}},
  \citenamefont {Sun},\ and\ \citenamefont {Duan}}]{PhysRevLett.123.140602}%
  \BibitemOpen
  \bibfield  {author} {\bibinfo {author} {\bibfnamefont {C.~B.}\ \bibnamefont
  {Da\ifmmode~\breve{g}\else \u{g}\fi{}}}, \bibinfo {author} {\bibfnamefont
  {K.}~\bibnamefont {Sun}}, \ and\ \bibinfo {author} {\bibfnamefont {L.-M.}\
  \bibnamefont {Duan}},\ }\bibfield  {title} {\enquote {\bibinfo {title}
  {Detection of quantum phases via out-of-time-order correlators},}\ }\href
  {\doibase 10.1103/PhysRevLett.123.140602} {\bibfield  {journal} {\bibinfo
  {journal} {Phys. Rev. Lett.}\ }\textbf {\bibinfo {volume} {123}},\ \bibinfo
  {pages} {140602} (\bibinfo {year} {2019})}\BibitemShut {NoStop}%
\bibitem [{\citenamefont {Maldacena}(1999)}]{Maldacena1999}%
  \BibitemOpen
  \bibfield  {author} {\bibinfo {author} {\bibfnamefont {J.}~\bibnamefont
  {Maldacena}},\ }\bibfield  {title} {\enquote {\bibinfo {title} {The large-n
  limit of superconformal field theories and supergravity},}\ }\href {\doibase
  10.1023/A:1026654312961} {\bibfield  {journal} {\bibinfo  {journal}
  {International Journal of Theoretical Physics}\ }\textbf {\bibinfo {volume}
  {38}},\ \bibinfo {pages} {1113} (\bibinfo {year} {1999})}\BibitemShut
  {NoStop}%
\bibitem [{\citenamefont {Sekino}\ and\ \citenamefont
  {Susskind}(2008)}]{Sekino_2008}%
  \BibitemOpen
  \bibfield  {author} {\bibinfo {author} {\bibfnamefont {Y.}~\bibnamefont
  {Sekino}}\ and\ \bibinfo {author} {\bibfnamefont {L.}~\bibnamefont
  {Susskind}},\ }\bibfield  {title} {\enquote {\bibinfo {title} {Fast
  scramblers},}\ }\href {\doibase 10.1088/1126-6708/2008/10/065} {\bibfield
  {journal} {\bibinfo  {journal} {Journal of High Energy Physics}\ }\textbf
  {\bibinfo {volume} {2008}},\ \bibinfo {pages} {065} (\bibinfo {year}
  {2008})}\BibitemShut {NoStop}%
\bibitem [{\citenamefont {Shenker}\ and\ \citenamefont
  {Stanford}(2014{\natexlab{a}})}]{Shenker2014}%
  \BibitemOpen
  \bibfield  {author} {\bibinfo {author} {\bibfnamefont {S.~H.}\ \bibnamefont
  {Shenker}}\ and\ \bibinfo {author} {\bibfnamefont {D.}~\bibnamefont
  {Stanford}},\ }\bibfield  {title} {\enquote {\bibinfo {title} {Black holes
  and the butterfly effect},}\ }\href {\doibase 10.1007/JHEP03(2014)067}
  {\bibfield  {journal} {\bibinfo  {journal} {Journal of High Energy Physics}\
  }\textbf {\bibinfo {volume} {2014}},\ \bibinfo {pages} {67} (\bibinfo {year}
  {2014}{\natexlab{a}})}\BibitemShut {NoStop}%
\bibitem [{\citenamefont {Shenker}\ and\ \citenamefont
  {Stanford}(2014{\natexlab{b}})}]{Shenker2014_2}%
  \BibitemOpen
  \bibfield  {author} {\bibinfo {author} {\bibfnamefont {S.~H.}\ \bibnamefont
  {Shenker}}\ and\ \bibinfo {author} {\bibfnamefont {D.}~\bibnamefont
  {Stanford}},\ }\bibfield  {title} {\enquote {\bibinfo {title} {Multiple
  shocks},}\ }\href {\doibase 10.1007/JHEP12(2014)046} {\bibfield  {journal}
  {\bibinfo  {journal} {Journal of High Energy Physics}\ }\textbf {\bibinfo
  {volume} {2014}},\ \bibinfo {pages} {46} (\bibinfo {year}
  {2014}{\natexlab{b}})}\BibitemShut {NoStop}%
\bibitem [{\citenamefont {Roberts}\ \emph {et~al.}(2015)\citenamefont
  {Roberts}, \citenamefont {Stanford},\ and\ \citenamefont
  {Susskind}}]{Roberts2015}%
  \BibitemOpen
  \bibfield  {author} {\bibinfo {author} {\bibfnamefont {D.~A.}\ \bibnamefont
  {Roberts}}, \bibinfo {author} {\bibfnamefont {D.}~\bibnamefont {Stanford}}, \
  and\ \bibinfo {author} {\bibfnamefont {L.}~\bibnamefont {Susskind}},\
  }\bibfield  {title} {\enquote {\bibinfo {title} {Localized shocks},}\ }\href
  {\doibase 10.1007/JHEP03(2015)051} {\bibfield  {journal} {\bibinfo  {journal}
  {Journal of High Energy Physics}\ }\textbf {\bibinfo {volume} {2015}},\
  \bibinfo {pages} {51} (\bibinfo {year} {2015})}\BibitemShut {NoStop}%
\bibitem [{\citenamefont {Shenker}\ and\ \citenamefont
  {Stanford}(2015)}]{Shenker2015}%
  \BibitemOpen
  \bibfield  {author} {\bibinfo {author} {\bibfnamefont {S.~H.}\ \bibnamefont
  {Shenker}}\ and\ \bibinfo {author} {\bibfnamefont {D.}~\bibnamefont
  {Stanford}},\ }\bibfield  {title} {\enquote {\bibinfo {title} {Stringy
  effects in scrambling},}\ }\href {\doibase 10.1007/JHEP05(2015)132}
  {\bibfield  {journal} {\bibinfo  {journal} {Journal of High Energy Physics}\
  }\textbf {\bibinfo {volume} {2015}},\ \bibinfo {pages} {132} (\bibinfo {year}
  {2015})}\BibitemShut {NoStop}%
\bibitem [{\citenamefont {Jensen}(2016)}]{PhysRevLett.117.111601}%
  \BibitemOpen
  \bibfield  {author} {\bibinfo {author} {\bibfnamefont {K.}~\bibnamefont
  {Jensen}},\ }\bibfield  {title} {\enquote {\bibinfo {title} {Chaos in
  ${\mathrm{ads}}_{2}$ holography},}\ }\href {\doibase
  10.1103/PhysRevLett.117.111601} {\bibfield  {journal} {\bibinfo  {journal}
  {Phys. Rev. Lett.}\ }\textbf {\bibinfo {volume} {117}},\ \bibinfo {pages}
  {111601} (\bibinfo {year} {2016})}\BibitemShut {NoStop}%
\bibitem [{\citenamefont {Maldacena}\ \emph {et~al.}(2017)\citenamefont
  {Maldacena}, \citenamefont {Stanford},\ and\ \citenamefont
  {Yang}}]{doi:10.1002/prop.201700034}%
  \BibitemOpen
  \bibfield  {author} {\bibinfo {author} {\bibfnamefont {J.}~\bibnamefont
  {Maldacena}}, \bibinfo {author} {\bibfnamefont {D.}~\bibnamefont {Stanford}},
  \ and\ \bibinfo {author} {\bibfnamefont {Z.}~\bibnamefont {Yang}},\
  }\bibfield  {title} {\enquote {\bibinfo {title} {Diving into traversable
  wormholes},}\ }\href {\doibase 10.1002/prop.201700034} {\bibfield  {journal}
  {\bibinfo  {journal} {Fortschritte der Physik}\ }\textbf {\bibinfo {volume}
  {65}},\ \bibinfo {pages} {1700034} (\bibinfo {year} {2017})}\BibitemShut
  {NoStop}%
\bibitem [{\citenamefont {Yoshida}\ and\ \citenamefont
  {Kitaev}(2017)}]{1710.03363}%
  \BibitemOpen
  \bibfield  {author} {\bibinfo {author} {\bibfnamefont {B.}~\bibnamefont
  {Yoshida}}\ and\ \bibinfo {author} {\bibfnamefont {A.}~\bibnamefont
  {Kitaev}},\ }\href@noop {} {\enquote {\bibinfo {title} {Efficient decoding
  for the hayden-preskill protocol},}\ } (\bibinfo {year} {2017}),\ \Eprint
  {http://arxiv.org/abs/arXiv:1710.03363} {arXiv:1710.03363} \BibitemShut
  {NoStop}%
\bibitem [{\citenamefont {G{\"a}rttner}\ \emph {et~al.}(2017)\citenamefont
  {G{\"a}rttner}, \citenamefont {Bohnet}, \citenamefont {Safavi-Naini},
  \citenamefont {Wall}, \citenamefont {Bollinger},\ and\ \citenamefont
  {Rey}}]{garttner2017measuring}%
  \BibitemOpen
  \bibfield  {author} {\bibinfo {author} {\bibfnamefont {M.}~\bibnamefont
  {G{\"a}rttner}}, \bibinfo {author} {\bibfnamefont {J.~G.}\ \bibnamefont
  {Bohnet}}, \bibinfo {author} {\bibfnamefont {A.}~\bibnamefont
  {Safavi-Naini}}, \bibinfo {author} {\bibfnamefont {M.~L.}\ \bibnamefont
  {Wall}}, \bibinfo {author} {\bibfnamefont {J.~J.}\ \bibnamefont {Bollinger}},
  \ and\ \bibinfo {author} {\bibfnamefont {A.~M.}\ \bibnamefont {Rey}},\
  }\bibfield  {title} {\enquote {\bibinfo {title} {Measuring out-of-time-order
  correlations and multiple quantum spectra in a trapped-ion quantum magnet},}\
  }\href {\doibase 10.1038/nphys4119} {\bibfield  {journal} {\bibinfo
  {journal} {Nature Physics}\ }\textbf {\bibinfo {volume} {13}},\ \bibinfo
  {pages} {781} (\bibinfo {year} {2017})}\BibitemShut {NoStop}%
\bibitem [{\citenamefont {Li}\ \emph {et~al.}(2017)\citenamefont {Li},
  \citenamefont {Fan}, \citenamefont {Wang}, \citenamefont {Ye}, \citenamefont
  {Zeng}, \citenamefont {Zhai}, \citenamefont {Peng},\ and\ \citenamefont
  {Du}}]{PhysRevX.7.031011}%
  \BibitemOpen
  \bibfield  {author} {\bibinfo {author} {\bibfnamefont {J.}~\bibnamefont
  {Li}}, \bibinfo {author} {\bibfnamefont {R.}~\bibnamefont {Fan}}, \bibinfo
  {author} {\bibfnamefont {H.}~\bibnamefont {Wang}}, \bibinfo {author}
  {\bibfnamefont {B.}~\bibnamefont {Ye}}, \bibinfo {author} {\bibfnamefont
  {B.}~\bibnamefont {Zeng}}, \bibinfo {author} {\bibfnamefont {H.}~\bibnamefont
  {Zhai}}, \bibinfo {author} {\bibfnamefont {X.}~\bibnamefont {Peng}}, \ and\
  \bibinfo {author} {\bibfnamefont {J.}~\bibnamefont {Du}},\ }\bibfield
  {title} {\enquote {\bibinfo {title} {Measuring out-of-time-order correlators
  on a nuclear magnetic resonance quantum simulator},}\ }\href {\doibase
  10.1103/PhysRevX.7.031011} {\bibfield  {journal} {\bibinfo  {journal} {Phys.
  Rev. X}\ }\textbf {\bibinfo {volume} {7}},\ \bibinfo {pages} {031011}
  (\bibinfo {year} {2017})}\BibitemShut {NoStop}%
\bibitem [{\citenamefont {Wei}\ \emph {et~al.}(2018)\citenamefont {Wei},
  \citenamefont {Ramanathan},\ and\ \citenamefont
  {Cappellaro}}]{PhysRevLett.120.070501}%
  \BibitemOpen
  \bibfield  {author} {\bibinfo {author} {\bibfnamefont {K.~X.}\ \bibnamefont
  {Wei}}, \bibinfo {author} {\bibfnamefont {C.}~\bibnamefont {Ramanathan}}, \
  and\ \bibinfo {author} {\bibfnamefont {P.}~\bibnamefont {Cappellaro}},\
  }\bibfield  {title} {\enquote {\bibinfo {title} {Exploring localization in
  nuclear spin chains},}\ }\href {\doibase 10.1103/PhysRevLett.120.070501}
  {\bibfield  {journal} {\bibinfo  {journal} {Phys. Rev. Lett.}\ }\textbf
  {\bibinfo {volume} {120}},\ \bibinfo {pages} {070501} (\bibinfo {year}
  {2018})}\BibitemShut {NoStop}%
\bibitem [{\citenamefont {Meier}\ \emph {et~al.}(2019)\citenamefont {Meier},
  \citenamefont {Ang'ong'a}, \citenamefont {An},\ and\ \citenamefont
  {Gadway}}]{PhysRevA.100.013623}%
  \BibitemOpen
  \bibfield  {author} {\bibinfo {author} {\bibfnamefont {E.~J.}\ \bibnamefont
  {Meier}}, \bibinfo {author} {\bibfnamefont {J.}~\bibnamefont {Ang'ong'a}},
  \bibinfo {author} {\bibfnamefont {F.~A.}\ \bibnamefont {An}}, \ and\ \bibinfo
  {author} {\bibfnamefont {B.}~\bibnamefont {Gadway}},\ }\bibfield  {title}
  {\enquote {\bibinfo {title} {Exploring quantum signatures of chaos on a
  floquet synthetic lattice},}\ }\href {\doibase 10.1103/PhysRevA.100.013623}
  {\bibfield  {journal} {\bibinfo  {journal} {Phys. Rev. A}\ }\textbf {\bibinfo
  {volume} {100}},\ \bibinfo {pages} {013623} (\bibinfo {year}
  {2019})}\BibitemShut {NoStop}%
\bibitem [{\citenamefont {Michalski}\ \emph {et~al.}(2013)\citenamefont
  {Michalski}, \citenamefont {Carbonell},\ and\ \citenamefont
  {Mitchell}}]{Michalski2013Machine}%
  \BibitemOpen
  \bibfield  {author} {\bibinfo {author} {\bibfnamefont {R.~S.}\ \bibnamefont
  {Michalski}}, \bibinfo {author} {\bibfnamefont {J.~G.}\ \bibnamefont
  {Carbonell}}, \ and\ \bibinfo {author} {\bibfnamefont {T.~M.}\ \bibnamefont
  {Mitchell}},\ }\href@noop {} {\emph {\bibinfo {title} {Machine learning: An
  artificial intelligence approach}}}\ (\bibinfo  {publisher} {Springer Science
  \& Business Media},\ \bibinfo {year} {2013})\BibitemShut {NoStop}%
\bibitem [{\citenamefont {LeCun}\ \emph {et~al.}(2015)\citenamefont {LeCun},
  \citenamefont {Bengio},\ and\ \citenamefont {Hinton}}]{LeCun2015Deep}%
  \BibitemOpen
  \bibfield  {author} {\bibinfo {author} {\bibfnamefont {Y.}~\bibnamefont
  {LeCun}}, \bibinfo {author} {\bibfnamefont {Y.}~\bibnamefont {Bengio}}, \
  and\ \bibinfo {author} {\bibfnamefont {G.}~\bibnamefont {Hinton}},\
  }\bibfield  {title} {\enquote {\bibinfo {title} {Deep learning},}\ }\href
  {\doibase 10.1038/nature14539} {\bibfield  {journal} {\bibinfo  {journal}
  {Nature}\ }\textbf {\bibinfo {volume} {521}},\ \bibinfo {pages} {436}
  (\bibinfo {year} {2015})}\BibitemShut {NoStop}%
\bibitem [{\citenamefont {Jordan}\ and\ \citenamefont
  {Mitchell}(2015)}]{Jordan2015Machine}%
  \BibitemOpen
  \bibfield  {author} {\bibinfo {author} {\bibfnamefont {M.}~\bibnamefont
  {Jordan}}\ and\ \bibinfo {author} {\bibfnamefont {T.}~\bibnamefont
  {Mitchell}},\ }\bibfield  {title} {\enquote {\bibinfo {title} {Machine
  learning: Trends, perspectives, and prospects},}\ }\href {\doibase
  10.1126/science.aaa8415} {\bibfield  {journal} {\bibinfo  {journal}
  {Science}\ }\textbf {\bibinfo {volume} {349}},\ \bibinfo {pages} {255}
  (\bibinfo {year} {2015})}\BibitemShut {NoStop}%
\bibitem [{\citenamefont {Kitaev}(2015)}]{kitaev2015otoc}%
  \BibitemOpen
  \bibfield  {author} {\bibinfo {author} {\bibfnamefont {A.}~\bibnamefont
  {Kitaev}},\ }\href@noop {} {\enquote {\bibinfo {title} {A simple model of
  quantum holography},}\ }\bibinfo {howpublished}
  {http://online.kitp.ucsb.edu/online/entangled15/kitaev/,
  http://online.kitp.ucsb.edu/online/entangled15/kitaev2/} (\bibinfo {year}
  {2015})\BibitemShut {NoStop}%
\bibitem [{\citenamefont {Polchinski}\ and\ \citenamefont
  {Rosenhaus}(2016)}]{Polchinski2016}%
  \BibitemOpen
  \bibfield  {author} {\bibinfo {author} {\bibfnamefont {J.}~\bibnamefont
  {Polchinski}}\ and\ \bibinfo {author} {\bibfnamefont {V.}~\bibnamefont
  {Rosenhaus}},\ }\bibfield  {title} {\enquote {\bibinfo {title} {The spectrum
  in the sachdev-ye-kitaev model},}\ }\href {\doibase 10.1007/JHEP04(2016)001}
  {\bibfield  {journal} {\bibinfo  {journal} {Journal of High Energy Physics}\
  }\textbf {\bibinfo {volume} {2016}},\ \bibinfo {pages} {1} (\bibinfo {year}
  {2016})}\BibitemShut {NoStop}%
\bibitem [{\citenamefont {Maldacena}\ and\ \citenamefont
  {Stanford}(2016)}]{PhysRevD.94.106002}%
  \BibitemOpen
  \bibfield  {author} {\bibinfo {author} {\bibfnamefont {J.}~\bibnamefont
  {Maldacena}}\ and\ \bibinfo {author} {\bibfnamefont {D.}~\bibnamefont
  {Stanford}},\ }\bibfield  {title} {\enquote {\bibinfo {title} {Remarks on the
  sachdev-ye-kitaev model},}\ }\href {\doibase 10.1103/PhysRevD.94.106002}
  {\bibfield  {journal} {\bibinfo  {journal} {Phys. Rev. D}\ }\textbf {\bibinfo
  {volume} {94}},\ \bibinfo {pages} {106002} (\bibinfo {year}
  {2016})}\BibitemShut {NoStop}%
\bibitem [{\citenamefont {McGinley}\ \emph {et~al.}(2019)\citenamefont
  {McGinley}, \citenamefont {Nunnenkamp},\ and\ \citenamefont
  {Knolle}}]{PhysRevLett.122.020603}%
  \BibitemOpen
  \bibfield  {author} {\bibinfo {author} {\bibfnamefont {M.}~\bibnamefont
  {McGinley}}, \bibinfo {author} {\bibfnamefont {A.}~\bibnamefont
  {Nunnenkamp}}, \ and\ \bibinfo {author} {\bibfnamefont {J.}~\bibnamefont
  {Knolle}},\ }\bibfield  {title} {\enquote {\bibinfo {title} {Slow growth of
  out-of-time-order correlators and entanglement entropy in integrable
  disordered systems},}\ }\href {\doibase 10.1103/PhysRevLett.122.020603}
  {\bibfield  {journal} {\bibinfo  {journal} {Phys. Rev. Lett.}\ }\textbf
  {\bibinfo {volume} {122}},\ \bibinfo {pages} {020603} (\bibinfo {year}
  {2019})}\BibitemShut {NoStop}%
\bibitem [{\citenamefont {Lin}\ and\ \citenamefont
  {Motrunich}(2018)}]{PhysRevB.97.144304}%
  \BibitemOpen
  \bibfield  {author} {\bibinfo {author} {\bibfnamefont {C.-J.}\ \bibnamefont
  {Lin}}\ and\ \bibinfo {author} {\bibfnamefont {O.~I.}\ \bibnamefont
  {Motrunich}},\ }\bibfield  {title} {\enquote {\bibinfo {title}
  {Out-of-time-ordered correlators in a quantum ising chain},}\ }\href
  {\doibase 10.1103/PhysRevB.97.144304} {\bibfield  {journal} {\bibinfo
  {journal} {Phys. Rev. B}\ }\textbf {\bibinfo {volume} {97}},\ \bibinfo
  {pages} {144304} (\bibinfo {year} {2018})}\BibitemShut {NoStop}%
\bibitem [{\citenamefont {Xu}\ and\ \citenamefont
  {Swingle}(2018)}]{1802.00801}%
  \BibitemOpen
  \bibfield  {author} {\bibinfo {author} {\bibfnamefont {S.}~\bibnamefont
  {Xu}}\ and\ \bibinfo {author} {\bibfnamefont {B.}~\bibnamefont {Swingle}},\
  }\href@noop {} {\enquote {\bibinfo {title} {Accessing scrambling using matrix
  product operators},}\ } (\bibinfo {year} {2018}),\ \Eprint
  {http://arxiv.org/abs/arXiv:1802.00801} {arXiv:1802.00801} \BibitemShut
  {NoStop}%
\bibitem [{\citenamefont {Schollw{\"o}ck}(2011)}]{Schollwock2011Density}%
  \BibitemOpen
  \bibfield  {author} {\bibinfo {author} {\bibfnamefont {U.}~\bibnamefont
  {Schollw{\"o}ck}},\ }\bibfield  {title} {\enquote {\bibinfo {title} {The
  density-matrix renormalization group in the age of matrix product states},}\
  }\href {\doibase 10.1016/j.aop.2010.09.012} {\bibfield  {journal} {\bibinfo
  {journal} {Ann. of Phys.}\ }\textbf {\bibinfo {volume} {326}},\ \bibinfo
  {pages} {96} (\bibinfo {year} {2011})}\BibitemShut {NoStop}%
\bibitem [{\citenamefont {Garc{\'\i}a-S{\'a}ez}\ and\ \citenamefont
  {Latorre}(2012)}]{garcia2011exact}%
  \BibitemOpen
  \bibfield  {author} {\bibinfo {author} {\bibfnamefont {A.}~\bibnamefont
  {Garc{\'\i}a-S{\'a}ez}}\ and\ \bibinfo {author} {\bibfnamefont {J.~I.}\
  \bibnamefont {Latorre}},\ }\bibfield  {title} {\enquote {\bibinfo {title} {An
  exact tensor network for the 3sat problem},}\ }\href@noop {} {\bibfield
  {journal} {\bibinfo  {journal} {Quantum Information and Computation}\
  }\textbf {\bibinfo {volume} {12}},\ \bibinfo {pages} {0283} (\bibinfo {year}
  {2012})}\BibitemShut {NoStop}%
\bibitem [{\citenamefont {Carleo}\ \emph {et~al.}(2019)\citenamefont {Carleo},
  \citenamefont {Cirac}, \citenamefont {Cranmer}, \citenamefont {Daudet},
  \citenamefont {Schuld}, \citenamefont {Tishby}, \citenamefont
  {Vogt-Maranto},\ and\ \citenamefont {Zdeborov{\'a}}}]{carleo2019machine}%
  \BibitemOpen
  \bibfield  {author} {\bibinfo {author} {\bibfnamefont {G.}~\bibnamefont
  {Carleo}}, \bibinfo {author} {\bibfnamefont {I.}~\bibnamefont {Cirac}},
  \bibinfo {author} {\bibfnamefont {K.}~\bibnamefont {Cranmer}}, \bibinfo
  {author} {\bibfnamefont {L.}~\bibnamefont {Daudet}}, \bibinfo {author}
  {\bibfnamefont {M.}~\bibnamefont {Schuld}}, \bibinfo {author} {\bibfnamefont
  {N.}~\bibnamefont {Tishby}}, \bibinfo {author} {\bibfnamefont
  {L.}~\bibnamefont {Vogt-Maranto}}, \ and\ \bibinfo {author} {\bibfnamefont
  {L.}~\bibnamefont {Zdeborov{\'a}}},\ }\bibfield  {title} {\enquote {\bibinfo
  {title} {Machine learning and the physical sciences},}\ }\href
  {https://arxiv.org/abs/1903.10563} {\bibfield  {journal} {\bibinfo  {journal}
  {arXiv:1903.10563}\ } (\bibinfo {year} {2019})}\BibitemShut {NoStop}%
\bibitem [{\citenamefont {Sarma}\ \emph {et~al.}(2019)\citenamefont {Sarma},
  \citenamefont {Deng},\ and\ \citenamefont {Duan}}]{Sarma2019Machine}%
  \BibitemOpen
  \bibfield  {author} {\bibinfo {author} {\bibfnamefont {S.~D.}\ \bibnamefont
  {Sarma}}, \bibinfo {author} {\bibfnamefont {D.-L.}\ \bibnamefont {Deng}}, \
  and\ \bibinfo {author} {\bibfnamefont {L.-M.}\ \bibnamefont {Duan}},\
  }\bibfield  {title} {\enquote {\bibinfo {title} {Machine learning meets
  quantum physics},}\ }\href {\doibase 10.1063/PT.3.4164} {\bibfield  {journal}
  {\bibinfo  {journal} {Physics Today}\ }\textbf {\bibinfo {volume} {72}},\
  \bibinfo {pages} {48} (\bibinfo {year} {2019})}\BibitemShut {NoStop}%
\bibitem [{\citenamefont {Carleo}\ and\ \citenamefont
  {Troyer}(2017)}]{Carleo2016Solving}%
  \BibitemOpen
  \bibfield  {author} {\bibinfo {author} {\bibfnamefont {G.}~\bibnamefont
  {Carleo}}\ and\ \bibinfo {author} {\bibfnamefont {M.}~\bibnamefont
  {Troyer}},\ }\bibfield  {title} {\enquote {\bibinfo {title} {Solving the
  quantum many-body problem with artificial neural networks},}\ }\href
  {\doibase 10.1126/science.aag2302} {\bibfield  {journal} {\bibinfo  {journal}
  {Science}\ }\textbf {\bibinfo {volume} {355}},\ \bibinfo {pages} {602}
  (\bibinfo {year} {2017})}\BibitemShut {NoStop}%
\bibitem [{\citenamefont {Deng}\ \emph
  {et~al.}(2017{\natexlab{a}})\citenamefont {Deng}, \citenamefont {Li},\ and\
  \citenamefont {Das~Sarma}}]{Deng2017Quantum}%
  \BibitemOpen
  \bibfield  {author} {\bibinfo {author} {\bibfnamefont {D.-L.}\ \bibnamefont
  {Deng}}, \bibinfo {author} {\bibfnamefont {X.}~\bibnamefont {Li}}, \ and\
  \bibinfo {author} {\bibfnamefont {S.}~\bibnamefont {Das~Sarma}},\ }\bibfield
  {title} {\enquote {\bibinfo {title} {Quantum entanglement in neural network
  states},}\ }\href {\doibase 10.1103/PhysRevX.7.021021} {\bibfield  {journal}
  {\bibinfo  {journal} {Phys. Rev. X}\ }\textbf {\bibinfo {volume} {7}},\
  \bibinfo {pages} {021021} (\bibinfo {year} {2017}{\natexlab{a}})}\BibitemShut
  {NoStop}%
\bibitem [{\citenamefont {Arsenault}\ \emph {et~al.}(2015)\citenamefont
  {Arsenault}, \citenamefont {von Lilienfeld},\ and\ \citenamefont
  {Millis}}]{Arsenault2015Machine}%
  \BibitemOpen
  \bibfield  {author} {\bibinfo {author} {\bibfnamefont {L.-F.}\ \bibnamefont
  {Arsenault}}, \bibinfo {author} {\bibfnamefont {O.~A.}\ \bibnamefont {von
  Lilienfeld}}, \ and\ \bibinfo {author} {\bibfnamefont {A.~J.}\ \bibnamefont
  {Millis}},\ }\bibfield  {title} {\enquote {\bibinfo {title} {Machine learning
  for many-body physics: efficient solution of dynamical mean-field theory},}\
  }\href {https://arxiv.org/abs/1506.08858} {\bibfield  {journal} {\bibinfo
  {journal} {arXiv:1506.08858}\ } (\bibinfo {year} {2015})}\BibitemShut
  {NoStop}%
\bibitem [{\citenamefont {Zhang}\ and\ \citenamefont
  {Kim}(2017)}]{Zhang2016Triangular}%
  \BibitemOpen
  \bibfield  {author} {\bibinfo {author} {\bibfnamefont {Y.}~\bibnamefont
  {Zhang}}\ and\ \bibinfo {author} {\bibfnamefont {E.-A.}\ \bibnamefont
  {Kim}},\ }\bibfield  {title} {\enquote {\bibinfo {title} {Quantum loop
  topography for machine learning},}\ }\href {\doibase
  10.1103/PhysRevLett.118.216401} {\bibfield  {journal} {\bibinfo  {journal}
  {Phys. Rev. Lett.}\ }\textbf {\bibinfo {volume} {118}},\ \bibinfo {pages}
  {216401} (\bibinfo {year} {2017})}\BibitemShut {NoStop}%
\bibitem [{\citenamefont {Carrasquilla}\ and\ \citenamefont
  {Melko}(2017)}]{Carrasquilla2017Machine}%
  \BibitemOpen
  \bibfield  {author} {\bibinfo {author} {\bibfnamefont {J.}~\bibnamefont
  {Carrasquilla}}\ and\ \bibinfo {author} {\bibfnamefont {R.~G.}\ \bibnamefont
  {Melko}},\ }\bibfield  {title} {\enquote {\bibinfo {title} {Machine learning
  phases of matter},}\ }\href {\doibase 10.1038/nphys4035} {\bibfield
  {journal} {\bibinfo  {journal} {Nat. Phys.}\ }\textbf {\bibinfo {volume}
  {13}},\ \bibinfo {pages} {431} (\bibinfo {year} {2017})}\BibitemShut
  {NoStop}%
\bibitem [{\citenamefont {van Nieuwenburg}\ \emph {et~al.}(2017)\citenamefont
  {van Nieuwenburg}, \citenamefont {Liu},\ and\ \citenamefont
  {Huber}}]{van2017Learning}%
  \BibitemOpen
  \bibfield  {author} {\bibinfo {author} {\bibfnamefont {E.~P.}\ \bibnamefont
  {van Nieuwenburg}}, \bibinfo {author} {\bibfnamefont {Y.-H.}\ \bibnamefont
  {Liu}}, \ and\ \bibinfo {author} {\bibfnamefont {S.~D.}\ \bibnamefont
  {Huber}},\ }\bibfield  {title} {\enquote {\bibinfo {title} {Learning phase
  transitions by confusion},}\ }\href {\doibase 10.1038/nphys4037} {\bibfield
  {journal} {\bibinfo  {journal} {Nat. Phys.}\ }\textbf {\bibinfo {volume}
  {13}},\ \bibinfo {pages} {435} (\bibinfo {year} {2017})}\BibitemShut
  {NoStop}%
\bibitem [{\citenamefont {Deng}\ \emph
  {et~al.}(2017{\natexlab{b}})\citenamefont {Deng}, \citenamefont {Li},\ and\
  \citenamefont {Das~Sarma}}]{Deng2017Machine}%
  \BibitemOpen
  \bibfield  {author} {\bibinfo {author} {\bibfnamefont {D.-L.}\ \bibnamefont
  {Deng}}, \bibinfo {author} {\bibfnamefont {X.}~\bibnamefont {Li}}, \ and\
  \bibinfo {author} {\bibfnamefont {S.}~\bibnamefont {Das~Sarma}},\ }\bibfield
  {title} {\enquote {\bibinfo {title} {Machine learning topological states},}\
  }\href {\doibase 10.1103/PhysRevB.96.195145} {\bibfield  {journal} {\bibinfo
  {journal} {Phys. Rev. B}\ }\textbf {\bibinfo {volume} {96}},\ \bibinfo
  {pages} {195145} (\bibinfo {year} {2017}{\natexlab{b}})}\BibitemShut
  {NoStop}%
\bibitem [{\citenamefont {Wang}(2016)}]{Wang2016Discovering}%
  \BibitemOpen
  \bibfield  {author} {\bibinfo {author} {\bibfnamefont {L.}~\bibnamefont
  {Wang}},\ }\bibfield  {title} {\enquote {\bibinfo {title} {Discovering phase
  transitions with unsupervised learning},}\ }\href {\doibase
  10.1103/PhysRevB.94.195105} {\bibfield  {journal} {\bibinfo  {journal} {Phys.
  Rev. B}\ }\textbf {\bibinfo {volume} {94}},\ \bibinfo {pages} {195105}
  (\bibinfo {year} {2016})}\BibitemShut {NoStop}%
\bibitem [{\citenamefont {Broecker}\ \emph
  {et~al.}(2017{\natexlab{a}})\citenamefont {Broecker}, \citenamefont
  {Carrasquilla}, \citenamefont {Melko},\ and\ \citenamefont
  {Trebst}}]{Broecker2017Machine}%
  \BibitemOpen
  \bibfield  {author} {\bibinfo {author} {\bibfnamefont {P.}~\bibnamefont
  {Broecker}}, \bibinfo {author} {\bibfnamefont {J.}~\bibnamefont
  {Carrasquilla}}, \bibinfo {author} {\bibfnamefont {R.~G.}\ \bibnamefont
  {Melko}}, \ and\ \bibinfo {author} {\bibfnamefont {S.}~\bibnamefont
  {Trebst}},\ }\bibfield  {title} {\enquote {\bibinfo {title} {Machine learning
  quantum phases of matter beyond the fermion sign problem},}\ }\href {\doibase
  10.1038/s41598-017-09098-0} {\bibfield  {journal} {\bibinfo  {journal} {Sci.
  Rep.}\ }\textbf {\bibinfo {volume} {7}} (\bibinfo {year}
  {2017}{\natexlab{a}}),\ 10.1038/s41598-017-09098-0}\BibitemShut {NoStop}%
\bibitem [{\citenamefont {Ch'ng}\ \emph {et~al.}(2017)\citenamefont {Ch'ng},
  \citenamefont {Carrasquilla}, \citenamefont {Melko},\ and\ \citenamefont
  {Khatami}}]{Chng2017Machine}%
  \BibitemOpen
  \bibfield  {author} {\bibinfo {author} {\bibfnamefont {K.}~\bibnamefont
  {Ch'ng}}, \bibinfo {author} {\bibfnamefont {J.}~\bibnamefont {Carrasquilla}},
  \bibinfo {author} {\bibfnamefont {R.~G.}\ \bibnamefont {Melko}}, \ and\
  \bibinfo {author} {\bibfnamefont {E.}~\bibnamefont {Khatami}},\ }\bibfield
  {title} {\enquote {\bibinfo {title} {Machine learning phases of strongly
  correlated fermions},}\ }\href {\doibase 10.1103/PhysRevX.7.031038}
  {\bibfield  {journal} {\bibinfo  {journal} {Phys. Rev. X}\ }\textbf {\bibinfo
  {volume} {7}},\ \bibinfo {pages} {031038} (\bibinfo {year}
  {2017})}\BibitemShut {NoStop}%
\bibitem [{\citenamefont {Zhang}\ \emph {et~al.}(2017)\citenamefont {Zhang},
  \citenamefont {Melko},\ and\ \citenamefont {Kim}}]{Zhang2017Machine}%
  \BibitemOpen
  \bibfield  {author} {\bibinfo {author} {\bibfnamefont {Y.}~\bibnamefont
  {Zhang}}, \bibinfo {author} {\bibfnamefont {R.~G.}\ \bibnamefont {Melko}}, \
  and\ \bibinfo {author} {\bibfnamefont {E.-A.}\ \bibnamefont {Kim}},\
  }\bibfield  {title} {\enquote {\bibinfo {title} {Machine learning $z_2$
  quantum spin liquids with quasiparticle statistics},}\ }\href {\doibase
  10.1103/PhysRevB.96.245119} {\bibfield  {journal} {\bibinfo  {journal} {Phys.
  Rev. B}\ }\textbf {\bibinfo {volume} {96}},\ \bibinfo {pages} {245119}
  (\bibinfo {year} {2017})}\BibitemShut {NoStop}%
\bibitem [{\citenamefont {Wetzel}(2017)}]{Wetzel2017Unsupervised}%
  \BibitemOpen
  \bibfield  {author} {\bibinfo {author} {\bibfnamefont {S.~J.}\ \bibnamefont
  {Wetzel}},\ }\bibfield  {title} {\enquote {\bibinfo {title} {Unsupervised
  learning of phase transitions: From principal component analysis to
  variational autoencoders},}\ }\href {\doibase 10.1103/PhysRevE.96.022140}
  {\bibfield  {journal} {\bibinfo  {journal} {Phys. Rev. E}\ }\textbf {\bibinfo
  {volume} {96}},\ \bibinfo {pages} {022140} (\bibinfo {year}
  {2017})}\BibitemShut {NoStop}%
\bibitem [{\citenamefont {Hu}\ \emph {et~al.}(2017)\citenamefont {Hu},
  \citenamefont {Singh},\ and\ \citenamefont {Scalettar}}]{Hu2017Discovering}%
  \BibitemOpen
  \bibfield  {author} {\bibinfo {author} {\bibfnamefont {W.}~\bibnamefont
  {Hu}}, \bibinfo {author} {\bibfnamefont {R.~R.~P.}\ \bibnamefont {Singh}}, \
  and\ \bibinfo {author} {\bibfnamefont {R.~T.}\ \bibnamefont {Scalettar}},\
  }\bibfield  {title} {\enquote {\bibinfo {title} {Discovering phases, phase
  transitions, and crossovers through unsupervised machine learning: A critical
  examination},}\ }\href {\doibase 10.1103/PhysRevE.95.062122} {\bibfield
  {journal} {\bibinfo  {journal} {Phys. Rev. E}\ }\textbf {\bibinfo {volume}
  {95}},\ \bibinfo {pages} {062122} (\bibinfo {year} {2017})}\BibitemShut
  {NoStop}%
\bibitem [{\citenamefont {Yoshioka}\ \emph {et~al.}(2018)\citenamefont
  {Yoshioka}, \citenamefont {Akagi},\ and\ \citenamefont
  {Katsura}}]{Yoshioka2017Learning}%
  \BibitemOpen
  \bibfield  {author} {\bibinfo {author} {\bibfnamefont {N.}~\bibnamefont
  {Yoshioka}}, \bibinfo {author} {\bibfnamefont {Y.}~\bibnamefont {Akagi}}, \
  and\ \bibinfo {author} {\bibfnamefont {H.}~\bibnamefont {Katsura}},\
  }\bibfield  {title} {\enquote {\bibinfo {title} {Learning disordered
  topological phases by statistical recovery of symmetry},}\ }\href {\doibase
  10.1103/PhysRevB.97.205110} {\bibfield  {journal} {\bibinfo  {journal} {Phys.
  Rev. B}\ }\textbf {\bibinfo {volume} {97}},\ \bibinfo {pages} {205110}
  (\bibinfo {year} {2018})}\BibitemShut {NoStop}%
\bibitem [{\citenamefont {Torlai}\ and\ \citenamefont
  {Melko}(2016)}]{Torlai2016Learning}%
  \BibitemOpen
  \bibfield  {author} {\bibinfo {author} {\bibfnamefont {G.}~\bibnamefont
  {Torlai}}\ and\ \bibinfo {author} {\bibfnamefont {R.~G.}\ \bibnamefont
  {Melko}},\ }\bibfield  {title} {\enquote {\bibinfo {title} {Learning
  thermodynamics with boltzmann machines},}\ }\href {\doibase
  10.1103/PhysRevB.94.165134} {\bibfield  {journal} {\bibinfo  {journal} {Phys.
  Rev. B}\ }\textbf {\bibinfo {volume} {94}},\ \bibinfo {pages} {165134}
  (\bibinfo {year} {2016})}\BibitemShut {NoStop}%
\bibitem [{\citenamefont {Aoki}\ and\ \citenamefont
  {Kobayashi}(2016)}]{Aoki2016restricted}%
  \BibitemOpen
  \bibfield  {author} {\bibinfo {author} {\bibfnamefont {K.-I.}\ \bibnamefont
  {Aoki}}\ and\ \bibinfo {author} {\bibfnamefont {T.}~\bibnamefont
  {Kobayashi}},\ }\bibfield  {title} {\enquote {\bibinfo {title} {Restricted
  boltzmann machines for the long range ising models},}\ }\href
  {http://www.worldscientific.com/doi/abs/10.1142/S0217984916504017} {\bibfield
   {journal} {\bibinfo  {journal} {Mod. Phys. Lett. B}\ ,\ \bibinfo {pages}
  {1650401}} (\bibinfo {year} {2016})}\BibitemShut {NoStop}%
\bibitem [{\citenamefont {You}\ \emph {et~al.}(2018)\citenamefont {You},
  \citenamefont {Yang},\ and\ \citenamefont {Qi}}]{You2017Machine}%
  \BibitemOpen
  \bibfield  {author} {\bibinfo {author} {\bibfnamefont {Y.-Z.}\ \bibnamefont
  {You}}, \bibinfo {author} {\bibfnamefont {Z.}~\bibnamefont {Yang}}, \ and\
  \bibinfo {author} {\bibfnamefont {X.-L.}\ \bibnamefont {Qi}},\ }\bibfield
  {title} {\enquote {\bibinfo {title} {Machine learning spatial geometry from
  entanglement features},}\ }\href {\doibase 10.1103/PhysRevB.97.045153}
  {\bibfield  {journal} {\bibinfo  {journal} {Phys. Rev. B}\ }\textbf {\bibinfo
  {volume} {97}},\ \bibinfo {pages} {045153} (\bibinfo {year}
  {2018})}\BibitemShut {NoStop}%
\bibitem [{\citenamefont {Torlai}\ \emph {et~al.}(2018)\citenamefont {Torlai},
  \citenamefont {Mazzola}, \citenamefont {Carrasquilla}, \citenamefont
  {Troyer}, \citenamefont {Melko},\ and\ \citenamefont
  {Carleo}}]{Torlai2018Neural}%
  \BibitemOpen
  \bibfield  {author} {\bibinfo {author} {\bibfnamefont {G.}~\bibnamefont
  {Torlai}}, \bibinfo {author} {\bibfnamefont {G.}~\bibnamefont {Mazzola}},
  \bibinfo {author} {\bibfnamefont {J.}~\bibnamefont {Carrasquilla}}, \bibinfo
  {author} {\bibfnamefont {M.}~\bibnamefont {Troyer}}, \bibinfo {author}
  {\bibfnamefont {R.}~\bibnamefont {Melko}}, \ and\ \bibinfo {author}
  {\bibfnamefont {G.}~\bibnamefont {Carleo}},\ }\bibfield  {title} {\enquote
  {\bibinfo {title} {Neural-network quantum state tomography},}\ }\href
  {\doibase 10.1038/s41567-018-0048-5} {\bibfield  {journal} {\bibinfo
  {journal} {Nat. Phys.}\ ,\ \bibinfo {pages} {1}} (\bibinfo {year}
  {2018})}\BibitemShut {NoStop}%
\bibitem [{\citenamefont {Pasquato}(2016)}]{Pasquato2016Detecting}%
  \BibitemOpen
  \bibfield  {author} {\bibinfo {author} {\bibfnamefont {M.}~\bibnamefont
  {Pasquato}},\ }\bibfield  {title} {\enquote {\bibinfo {title} {Detecting
  intermediate mass black holes in globular clusters with machine learning},}\
  }\href {https://arxiv.org/abs/1606.08548} {\bibfield  {journal} {\bibinfo
  {journal} {arXiv:1606.08548}\ } (\bibinfo {year} {2016})}\BibitemShut
  {NoStop}%
\bibitem [{\citenamefont {Hezaveh}\ \emph {et~al.}(2017)\citenamefont
  {Hezaveh}, \citenamefont {Perreault~Levasseur},\ and\ \citenamefont
  {Marshall}}]{Hezaveh2017Fast}%
  \BibitemOpen
  \bibfield  {author} {\bibinfo {author} {\bibfnamefont {Y.~D.}\ \bibnamefont
  {Hezaveh}}, \bibinfo {author} {\bibfnamefont {L.}~\bibnamefont
  {Perreault~Levasseur}}, \ and\ \bibinfo {author} {\bibfnamefont {P.~J.}\
  \bibnamefont {Marshall}},\ }\bibfield  {title} {\enquote {\bibinfo {title}
  {Fast automated analysis of strong gravitational lenses with convolutional
  neural networks},}\ }\href {\doibase 10.1038/nature23463} {\bibfield
  {journal} {\bibinfo  {journal} {Nature}\ }\textbf {\bibinfo {volume} {548}},\
  \bibinfo {pages} {555} (\bibinfo {year} {2017})}\BibitemShut {NoStop}%
\bibitem [{\citenamefont {Biswas}\ \emph {et~al.}(2013)\citenamefont {Biswas},
  \citenamefont {Blackburn}, \citenamefont {Cao}, \citenamefont {Essick},
  \citenamefont {Hodge}, \citenamefont {Katsavounidis}, \citenamefont {Kim},
  \citenamefont {Kim}, \citenamefont {Le~Bigot}, \citenamefont {Lee},
  \citenamefont {Oh}, \citenamefont {Oh}, \citenamefont {Son}, \citenamefont
  {Tao}, \citenamefont {Vaulin},\ and\ \citenamefont
  {Wang}}]{Rahul2013Application}%
  \BibitemOpen
  \bibfield  {author} {\bibinfo {author} {\bibfnamefont {R.}~\bibnamefont
  {Biswas}}, \bibinfo {author} {\bibfnamefont {L.}~\bibnamefont {Blackburn}},
  \bibinfo {author} {\bibfnamefont {J.}~\bibnamefont {Cao}}, \bibinfo {author}
  {\bibfnamefont {R.}~\bibnamefont {Essick}}, \bibinfo {author} {\bibfnamefont
  {K.~A.}\ \bibnamefont {Hodge}}, \bibinfo {author} {\bibfnamefont
  {E.}~\bibnamefont {Katsavounidis}}, \bibinfo {author} {\bibfnamefont
  {K.}~\bibnamefont {Kim}}, \bibinfo {author} {\bibfnamefont {Y.-M.}\
  \bibnamefont {Kim}}, \bibinfo {author} {\bibfnamefont {E.-O.}\ \bibnamefont
  {Le~Bigot}}, \bibinfo {author} {\bibfnamefont {C.-H.}\ \bibnamefont {Lee}},
  \bibinfo {author} {\bibfnamefont {J.~J.}\ \bibnamefont {Oh}}, \bibinfo
  {author} {\bibfnamefont {S.~H.}\ \bibnamefont {Oh}}, \bibinfo {author}
  {\bibfnamefont {E.~J.}\ \bibnamefont {Son}}, \bibinfo {author} {\bibfnamefont
  {Y.}~\bibnamefont {Tao}}, \bibinfo {author} {\bibfnamefont {R.}~\bibnamefont
  {Vaulin}}, \ and\ \bibinfo {author} {\bibfnamefont {X.}~\bibnamefont
  {Wang}},\ }\bibfield  {title} {\enquote {\bibinfo {title} {Application of
  machine learning algorithms to the study of noise artifacts in
  gravitational-wave data},}\ }\href {\doibase 10.1103/PhysRevD.88.062003}
  {\bibfield  {journal} {\bibinfo  {journal} {Phys. Rev. D}\ }\textbf {\bibinfo
  {volume} {88}},\ \bibinfo {pages} {062003} (\bibinfo {year}
  {2013})}\BibitemShut {NoStop}%
\bibitem [{\citenamefont {Abbott~{\it et al.}}(2016)}]{Abbott2016Observation}%
  \BibitemOpen
  \bibfield  {author} {\bibinfo {author} {\bibfnamefont {B.~P.}\ \bibnamefont
  {Abbott~{\it et al.}}} (\bibinfo {collaboration} {LIGO Scientific
  Collaboration and Virgo Collaboration}),\ }\bibfield  {title} {\enquote
  {\bibinfo {title} {Observation of gravitational waves from a binary black
  hole merger},}\ }\href {\doibase 10.1103/PhysRevLett.116.061102} {\bibfield
  {journal} {\bibinfo  {journal} {Phys. Rev. Lett.}\ }\textbf {\bibinfo
  {volume} {116}},\ \bibinfo {pages} {061102} (\bibinfo {year}
  {2016})}\BibitemShut {NoStop}%
\bibitem [{\citenamefont {Kalinin}\ \emph {et~al.}(2015)\citenamefont
  {Kalinin}, \citenamefont {Sumpter},\ and\ \citenamefont
  {Archibald}}]{Kalinin2015Big}%
  \BibitemOpen
  \bibfield  {author} {\bibinfo {author} {\bibfnamefont {S.~V.}\ \bibnamefont
  {Kalinin}}, \bibinfo {author} {\bibfnamefont {B.~G.}\ \bibnamefont
  {Sumpter}}, \ and\ \bibinfo {author} {\bibfnamefont {R.~K.}\ \bibnamefont
  {Archibald}},\ }\bibfield  {title} {\enquote {\bibinfo {title}
  {Big-deep-smart data in imaging for guiding materials design},}\ }\href
  {\doibase 10.1038/nmat4395} {\bibfield  {journal} {\bibinfo  {journal} {Nat.
  Mater.}\ }\textbf {\bibinfo {volume} {14}},\ \bibinfo {pages} {973} (\bibinfo
  {year} {2015})}\BibitemShut {NoStop}%
\bibitem [{\citenamefont {Schoenholz}\ \emph {et~al.}(2016)\citenamefont
  {Schoenholz}, \citenamefont {Cubuk}, \citenamefont {Sussman}, \citenamefont
  {Kaxiras},\ and\ \citenamefont {Liu}}]{Schoenholz2016Structural}%
  \BibitemOpen
  \bibfield  {author} {\bibinfo {author} {\bibfnamefont {S.~S.}\ \bibnamefont
  {Schoenholz}}, \bibinfo {author} {\bibfnamefont {E.~D.}\ \bibnamefont
  {Cubuk}}, \bibinfo {author} {\bibfnamefont {D.~M.}\ \bibnamefont {Sussman}},
  \bibinfo {author} {\bibfnamefont {E.}~\bibnamefont {Kaxiras}}, \ and\
  \bibinfo {author} {\bibfnamefont {A.~J.}\ \bibnamefont {Liu}},\ }\bibfield
  {title} {\enquote {\bibinfo {title} {A structural approach to relaxation in
  glassy liquids},}\ }\href {\doibase 10.1038/nphys3644} {\bibfield  {journal}
  {\bibinfo  {journal} {Nat. Phys.}\ }\textbf {\bibinfo {volume} {12}},\
  \bibinfo {pages} {469} (\bibinfo {year} {2016})}\BibitemShut {NoStop}%
\bibitem [{\citenamefont {Liu}\ \emph {et~al.}(2017)\citenamefont {Liu},
  \citenamefont {Qi}, \citenamefont {Meng},\ and\ \citenamefont
  {Fu}}]{Liu2017Self}%
  \BibitemOpen
  \bibfield  {author} {\bibinfo {author} {\bibfnamefont {J.}~\bibnamefont
  {Liu}}, \bibinfo {author} {\bibfnamefont {Y.}~\bibnamefont {Qi}}, \bibinfo
  {author} {\bibfnamefont {Z.~Y.}\ \bibnamefont {Meng}}, \ and\ \bibinfo
  {author} {\bibfnamefont {L.}~\bibnamefont {Fu}},\ }\bibfield  {title}
  {\enquote {\bibinfo {title} {Self-learning monte carlo method},}\ }\href
  {\doibase 10.1103/PhysRevB.95.041101} {\bibfield  {journal} {\bibinfo
  {journal} {Phys. Rev. B}\ }\textbf {\bibinfo {volume} {95}},\ \bibinfo
  {pages} {041101} (\bibinfo {year} {2017})}\BibitemShut {NoStop}%
\bibitem [{\citenamefont {Huang}\ and\ \citenamefont
  {Wang}(2017)}]{Huang2017Accelerated}%
  \BibitemOpen
  \bibfield  {author} {\bibinfo {author} {\bibfnamefont {L.}~\bibnamefont
  {Huang}}\ and\ \bibinfo {author} {\bibfnamefont {L.}~\bibnamefont {Wang}},\
  }\bibfield  {title} {\enquote {\bibinfo {title} {Accelerated monte carlo
  simulations with restricted boltzmann machines},}\ }\href {\doibase
  10.1103/PhysRevB.95.035105} {\bibfield  {journal} {\bibinfo  {journal} {Phys.
  Rev. B}\ }\textbf {\bibinfo {volume} {95}},\ \bibinfo {pages} {035105}
  (\bibinfo {year} {2017})}\BibitemShut {NoStop}%
\bibitem [{\citenamefont {Torlai}\ and\ \citenamefont
  {Melko}(2017)}]{Torlai2017Neural}%
  \BibitemOpen
  \bibfield  {author} {\bibinfo {author} {\bibfnamefont {G.}~\bibnamefont
  {Torlai}}\ and\ \bibinfo {author} {\bibfnamefont {R.~G.}\ \bibnamefont
  {Melko}},\ }\bibfield  {title} {\enquote {\bibinfo {title} {Neural decoder
  for topological codes},}\ }\href {\doibase 10.1103/PhysRevLett.119.030501}
  {\bibfield  {journal} {\bibinfo  {journal} {Phys. Rev. Lett.}\ }\textbf
  {\bibinfo {volume} {119}},\ \bibinfo {pages} {030501} (\bibinfo {year}
  {2017})}\BibitemShut {NoStop}%
\bibitem [{\citenamefont {Gao}\ and\ \citenamefont
  {Duan}(2017)}]{Gao2017Efficient}%
  \BibitemOpen
  \bibfield  {author} {\bibinfo {author} {\bibfnamefont {X.}~\bibnamefont
  {Gao}}\ and\ \bibinfo {author} {\bibfnamefont {L.-M.}\ \bibnamefont {Duan}},\
  }\bibfield  {title} {\enquote {\bibinfo {title} {Efficient representation of
  quantum many-body states with deep neural networks},}\ }\href {\doibase
  10.1038/s41467-017-00705-2} {\bibfield  {journal} {\bibinfo  {journal} {Nat.
  Commu.}\ ,\ \bibinfo {pages} {662}} (\bibinfo {year} {2017})}\BibitemShut
  {NoStop}%
\bibitem [{\citenamefont {Chen}\ \emph {et~al.}(2018)\citenamefont {Chen},
  \citenamefont {Cheng}, \citenamefont {Xie}, \citenamefont {Wang},\ and\
  \citenamefont {Xiang}}]{Chen2017Equivalence}%
  \BibitemOpen
  \bibfield  {author} {\bibinfo {author} {\bibfnamefont {J.}~\bibnamefont
  {Chen}}, \bibinfo {author} {\bibfnamefont {S.}~\bibnamefont {Cheng}},
  \bibinfo {author} {\bibfnamefont {H.}~\bibnamefont {Xie}}, \bibinfo {author}
  {\bibfnamefont {L.}~\bibnamefont {Wang}}, \ and\ \bibinfo {author}
  {\bibfnamefont {T.}~\bibnamefont {Xiang}},\ }\bibfield  {title} {\enquote
  {\bibinfo {title} {Equivalence of restricted boltzmann machines and tensor
  network states},}\ }\href {\doibase 10.1103/PhysRevB.97.085104} {\bibfield
  {journal} {\bibinfo  {journal} {Phys. Rev. B}\ }\textbf {\bibinfo {volume}
  {97}},\ \bibinfo {pages} {085104} (\bibinfo {year} {2018})}\BibitemShut
  {NoStop}%
\bibitem [{\citenamefont {Huang}\ and\ \citenamefont
  {Moore}(2017)}]{Huang2017Neural}%
  \BibitemOpen
  \bibfield  {author} {\bibinfo {author} {\bibfnamefont {Y.}~\bibnamefont
  {Huang}}\ and\ \bibinfo {author} {\bibfnamefont {J.~E.}\ \bibnamefont
  {Moore}},\ }\bibfield  {title} {\enquote {\bibinfo {title} {Neural network
  representation of tensor network and chiral states},}\ }\href
  {https://arxiv.org/abs/1701.06246} {\bibfield  {journal} {\bibinfo  {journal}
  {arXiv:1701.06246}\ } (\bibinfo {year} {2017})}\BibitemShut {NoStop}%
\bibitem [{\citenamefont {Schindler}\ \emph {et~al.}(2017)\citenamefont
  {Schindler}, \citenamefont {Regnault},\ and\ \citenamefont
  {Neupert}}]{Schindler2017Probing}%
  \BibitemOpen
  \bibfield  {author} {\bibinfo {author} {\bibfnamefont {F.}~\bibnamefont
  {Schindler}}, \bibinfo {author} {\bibfnamefont {N.}~\bibnamefont {Regnault}},
  \ and\ \bibinfo {author} {\bibfnamefont {T.}~\bibnamefont {Neupert}},\
  }\bibfield  {title} {\enquote {\bibinfo {title} {Probing many-body
  localization with neural networks},}\ }\href {\doibase
  10.1103/PhysRevB.95.245134} {\bibfield  {journal} {\bibinfo  {journal} {Phys.
  Rev. B}\ }\textbf {\bibinfo {volume} {95}},\ \bibinfo {pages} {245134}
  (\bibinfo {year} {2017})}\BibitemShut {NoStop}%
\bibitem [{\citenamefont {Cai}\ and\ \citenamefont
  {Liu}(2018)}]{Cai2017Approximating}%
  \BibitemOpen
  \bibfield  {author} {\bibinfo {author} {\bibfnamefont {Z.}~\bibnamefont
  {Cai}}\ and\ \bibinfo {author} {\bibfnamefont {J.}~\bibnamefont {Liu}},\
  }\bibfield  {title} {\enquote {\bibinfo {title} {Approximating quantum
  many-body wave functions using artificial neural networks},}\ }\href
  {\doibase 10.1103/PhysRevB.97.035116} {\bibfield  {journal} {\bibinfo
  {journal} {Phys. Rev. B}\ }\textbf {\bibinfo {volume} {97}},\ \bibinfo
  {pages} {035116} (\bibinfo {year} {2018})}\BibitemShut {NoStop}%
\bibitem [{\citenamefont {Broecker}\ \emph
  {et~al.}(2017{\natexlab{b}})\citenamefont {Broecker}, \citenamefont
  {Assaad},\ and\ \citenamefont {Trebst}}]{Broecker2017Quantum}%
  \BibitemOpen
  \bibfield  {author} {\bibinfo {author} {\bibfnamefont {P.}~\bibnamefont
  {Broecker}}, \bibinfo {author} {\bibfnamefont {F.~F.}\ \bibnamefont
  {Assaad}}, \ and\ \bibinfo {author} {\bibfnamefont {S.}~\bibnamefont
  {Trebst}},\ }\bibfield  {title} {\enquote {\bibinfo {title} {Quantum phase
  recognition via unsupervised machine learning},}\ }\href
  {https://arxiv.org/abs/1707.00663} {\bibfield  {journal} {\bibinfo  {journal}
  {arXiv:1707.00663}\ } (\bibinfo {year} {2017}{\natexlab{b}})}\BibitemShut
  {NoStop}%
\bibitem [{\citenamefont {Nomura}\ \emph {et~al.}(2017)\citenamefont {Nomura},
  \citenamefont {Darmawan}, \citenamefont {Yamaji},\ and\ \citenamefont
  {Imada}}]{Nomura2017Restricted}%
  \BibitemOpen
  \bibfield  {author} {\bibinfo {author} {\bibfnamefont {Y.}~\bibnamefont
  {Nomura}}, \bibinfo {author} {\bibfnamefont {A.~S.}\ \bibnamefont
  {Darmawan}}, \bibinfo {author} {\bibfnamefont {Y.}~\bibnamefont {Yamaji}}, \
  and\ \bibinfo {author} {\bibfnamefont {M.}~\bibnamefont {Imada}},\ }\bibfield
   {title} {\enquote {\bibinfo {title} {Restricted boltzmann machine learning
  for solving strongly correlated quantum systems},}\ }\href {\doibase
  10.1103/PhysRevB.96.205152} {\bibfield  {journal} {\bibinfo  {journal} {Phys.
  Rev. B}\ }\textbf {\bibinfo {volume} {96}},\ \bibinfo {pages} {205152}
  (\bibinfo {year} {2017})}\BibitemShut {NoStop}%
\bibitem [{\citenamefont {Biamonte}\ \emph {et~al.}(2017)\citenamefont
  {Biamonte}, \citenamefont {Wittek}, \citenamefont {Pancotti}, \citenamefont
  {Rebentrost}, \citenamefont {Wiebe},\ and\ \citenamefont
  {Lloyd}}]{Biamonte2017Quantum}%
  \BibitemOpen
  \bibfield  {author} {\bibinfo {author} {\bibfnamefont {J.}~\bibnamefont
  {Biamonte}}, \bibinfo {author} {\bibfnamefont {P.}~\bibnamefont {Wittek}},
  \bibinfo {author} {\bibfnamefont {N.}~\bibnamefont {Pancotti}}, \bibinfo
  {author} {\bibfnamefont {P.}~\bibnamefont {Rebentrost}}, \bibinfo {author}
  {\bibfnamefont {N.}~\bibnamefont {Wiebe}}, \ and\ \bibinfo {author}
  {\bibfnamefont {S.}~\bibnamefont {Lloyd}},\ }\bibfield  {title} {\enquote
  {\bibinfo {title} {Quantum machine learning},}\ }\href {\doibase
  10.1038/nature23474} {\bibfield  {journal} {\bibinfo  {journal} {Nature}\
  }\textbf {\bibinfo {volume} {549}},\ \bibinfo {pages} {195} (\bibinfo {year}
  {2017})}\BibitemShut {NoStop}%
\bibitem [{\citenamefont {Lu}\ \emph {et~al.}(2018)\citenamefont {Lu},
  \citenamefont {Huang}, \citenamefont {Li}, \citenamefont {Li}, \citenamefont
  {Chen}, \citenamefont {Lu}, \citenamefont {Ji}, \citenamefont {Shen},
  \citenamefont {Zhou},\ and\ \citenamefont {Zeng}}]{Lu2017Separability}%
  \BibitemOpen
  \bibfield  {author} {\bibinfo {author} {\bibfnamefont {S.}~\bibnamefont
  {Lu}}, \bibinfo {author} {\bibfnamefont {S.}~\bibnamefont {Huang}}, \bibinfo
  {author} {\bibfnamefont {K.}~\bibnamefont {Li}}, \bibinfo {author}
  {\bibfnamefont {J.}~\bibnamefont {Li}}, \bibinfo {author} {\bibfnamefont
  {J.}~\bibnamefont {Chen}}, \bibinfo {author} {\bibfnamefont {D.}~\bibnamefont
  {Lu}}, \bibinfo {author} {\bibfnamefont {Z.}~\bibnamefont {Ji}}, \bibinfo
  {author} {\bibfnamefont {Y.}~\bibnamefont {Shen}}, \bibinfo {author}
  {\bibfnamefont {D.}~\bibnamefont {Zhou}}, \ and\ \bibinfo {author}
  {\bibfnamefont {B.}~\bibnamefont {Zeng}},\ }\bibfield  {title} {\enquote
  {\bibinfo {title} {Separability-entanglement classifier via machine
  learning},}\ }\href {\doibase 10.1103/PhysRevA.98.012315} {\bibfield
  {journal} {\bibinfo  {journal} {Phys. Rev. A}\ }\textbf {\bibinfo {volume}
  {98}},\ \bibinfo {pages} {012315} (\bibinfo {year} {2018})}\BibitemShut
  {NoStop}%
\bibitem [{\citenamefont {Weinstein}(2017)}]{Weinstein2017Learning}%
  \BibitemOpen
  \bibfield  {author} {\bibinfo {author} {\bibfnamefont {S.}~\bibnamefont
  {Weinstein}},\ }\bibfield  {title} {\enquote {\bibinfo {title} {Learning the
  einstein-podolsky-rosen correlations on a restricted boltzmann machine},}\
  }\href {https://arxiv.org/abs/1707.03114} {\bibfield  {journal} {\bibinfo
  {journal} {arXiv preprint arXiv:1707.03114}\ } (\bibinfo {year}
  {2017})}\BibitemShut {NoStop}%
\bibitem [{\citenamefont {Saito}(2017)}]{Saito2017Solving}%
  \BibitemOpen
  \bibfield  {author} {\bibinfo {author} {\bibfnamefont {H.}~\bibnamefont
  {Saito}},\ }\bibfield  {title} {\enquote {\bibinfo {title} {Solving the
  bose--hubbard model with machine learning},}\ }\href {\doibase
  10.7566/JPSJ.86.093001} {\bibfield  {journal} {\bibinfo  {journal} {J. Phys.
  Soc. Jpn.}\ }\textbf {\bibinfo {volume} {86}},\ \bibinfo {pages} {093001}
  (\bibinfo {year} {2017})}\BibitemShut {NoStop}%
\bibitem [{\citenamefont {Saito}\ and\ \citenamefont
  {Kato}(2017)}]{Saito2017Machine}%
  \BibitemOpen
  \bibfield  {author} {\bibinfo {author} {\bibfnamefont {H.}~\bibnamefont
  {Saito}}\ and\ \bibinfo {author} {\bibfnamefont {M.}~\bibnamefont {Kato}},\
  }\bibfield  {title} {\enquote {\bibinfo {title} {Machine learning technique
  to find quantum many-body ground states of bosons on a lattice},}\ }\href
  {\doibase 10.7566/JPSJ.87.014001} {\bibfield  {journal} {\bibinfo  {journal}
  {J. Phys. Soc. Jpn.}\ }\textbf {\bibinfo {volume} {87}},\ \bibinfo {pages}
  {014001} (\bibinfo {year} {2017})}\BibitemShut {NoStop}%
\bibitem [{\citenamefont {Schmitt}\ and\ \citenamefont
  {Heyl}(2018)}]{Schmitt2017Quantum}%
  \BibitemOpen
  \bibfield  {author} {\bibinfo {author} {\bibfnamefont {M.}~\bibnamefont
  {Schmitt}}\ and\ \bibinfo {author} {\bibfnamefont {M.}~\bibnamefont {Heyl}},\
  }\bibfield  {title} {\enquote {\bibinfo {title} {Quantum dynamics in
  transverse-field ising models from classical networks},}\ }\href {\doibase
  10.21468/SciPostPhys.4.2.013} {\bibfield  {journal} {\bibinfo  {journal}
  {SciPost Phys.}\ }\textbf {\bibinfo {volume} {4}},\ \bibinfo {pages} {013}
  (\bibinfo {year} {2018})}\BibitemShut {NoStop}%
\bibitem [{\citenamefont {Deng}(2018)}]{Deng2017MachineBN}%
  \BibitemOpen
  \bibfield  {author} {\bibinfo {author} {\bibfnamefont {D.-L.}\ \bibnamefont
  {Deng}},\ }\bibfield  {title} {\enquote {\bibinfo {title} {Machine learning
  detection of bell nonlocality in quantum many-body systems},}\ }\href
  {\doibase 10.1103/PhysRevLett.120.240402} {\bibfield  {journal} {\bibinfo
  {journal} {Phys. Rev. Lett.}\ }\textbf {\bibinfo {volume} {120}},\ \bibinfo
  {pages} {240402} (\bibinfo {year} {2018})}\BibitemShut {NoStop}%
\bibitem [{\citenamefont {Hsu}\ \emph {et~al.}(2018)\citenamefont {Hsu},
  \citenamefont {Li}, \citenamefont {Deng},\ and\ \citenamefont
  {Das~Sarma}}]{Hsu2018Machine}%
  \BibitemOpen
  \bibfield  {author} {\bibinfo {author} {\bibfnamefont {Y.-T.}\ \bibnamefont
  {Hsu}}, \bibinfo {author} {\bibfnamefont {X.}~\bibnamefont {Li}}, \bibinfo
  {author} {\bibfnamefont {D.-L.}\ \bibnamefont {Deng}}, \ and\ \bibinfo
  {author} {\bibfnamefont {S.}~\bibnamefont {Das~Sarma}},\ }\bibfield  {title}
  {\enquote {\bibinfo {title} {Machine learning many-body localization: Search
  for the elusive nonergodic metal},}\ }\href {\doibase
  10.1103/PhysRevLett.121.245701} {\bibfield  {journal} {\bibinfo  {journal}
  {Phys. Rev. Lett.}\ }\textbf {\bibinfo {volume} {121}},\ \bibinfo {pages}
  {245701} (\bibinfo {year} {2018})}\BibitemShut {NoStop}%
\bibitem [{\citenamefont {Hinton}\ and\ \citenamefont
  {Salakhutdinov}(2006)}]{Hinton2006Reducing}%
  \BibitemOpen
  \bibfield  {author} {\bibinfo {author} {\bibfnamefont {G.~E.}\ \bibnamefont
  {Hinton}}\ and\ \bibinfo {author} {\bibfnamefont {R.~R.}\ \bibnamefont
  {Salakhutdinov}},\ }\bibfield  {title} {\enquote {\bibinfo {title} {Reducing
  the dimensionality of data with neural networks},}\ }\href {\doibase
  10.1126/science.1127647} {\bibfield  {journal} {\bibinfo  {journal}
  {Science}\ }\textbf {\bibinfo {volume} {313}},\ \bibinfo {pages} {504}
  (\bibinfo {year} {2006})}\BibitemShut {NoStop}%
\bibitem [{\citenamefont {Salakhutdinov}\ \emph {et~al.}(2007)\citenamefont
  {Salakhutdinov}, \citenamefont {Mnih},\ and\ \citenamefont
  {Hinton}}]{Salakhutdinov2007Restricted}%
  \BibitemOpen
  \bibfield  {author} {\bibinfo {author} {\bibfnamefont {R.}~\bibnamefont
  {Salakhutdinov}}, \bibinfo {author} {\bibfnamefont {A.}~\bibnamefont {Mnih}},
  \ and\ \bibinfo {author} {\bibfnamefont {G.}~\bibnamefont {Hinton}},\
  }\bibfield  {title} {\enquote {\bibinfo {title} {Restricted boltzmann
  machines for collaborative filtering},}\ }in\ \href
  {http://dl.acm.org/citation.cfm?id=1273596} {\emph {\bibinfo {booktitle}
  {Proceedings of the 24th international conference on Machine learning}}}\
  (\bibinfo {organization} {ACM},\ \bibinfo {year} {2007})\ pp.\ \bibinfo
  {pages} {791--798}\BibitemShut {NoStop}%
\bibitem [{\citenamefont {Larochelle}\ and\ \citenamefont
  {Bengio}(2008)}]{Larochelle2008Classification}%
  \BibitemOpen
  \bibfield  {author} {\bibinfo {author} {\bibfnamefont {H.}~\bibnamefont
  {Larochelle}}\ and\ \bibinfo {author} {\bibfnamefont {Y.}~\bibnamefont
  {Bengio}},\ }\bibfield  {title} {\enquote {\bibinfo {title} {Classification
  using discriminative restricted boltzmann machines},}\ }in\ \href
  {http://dl.acm.org/citation.cfm?id=1390224} {\emph {\bibinfo {booktitle}
  {Proceedings of the 25th international conference on Machine learning}}}\
  (\bibinfo {organization} {ACM},\ \bibinfo {year} {2008})\ pp.\ \bibinfo
  {pages} {536--543}\BibitemShut {NoStop}%
\bibitem [{\citenamefont {Swingle}(2018)}]{swingle2018unscrambling}%
  \BibitemOpen
  \bibfield  {author} {\bibinfo {author} {\bibfnamefont {B.}~\bibnamefont
  {Swingle}},\ }\bibfield  {title} {\enquote {\bibinfo {title} {Unscrambling
  the physics of out-of-time-order correlators},}\ }\href {\doibase
  10.1038/s41567-018-0295-5} {\bibfield  {journal} {\bibinfo  {journal} {Nature
  Physics}\ }\textbf {\bibinfo {volume} {14}},\ \bibinfo {pages} {988}
  (\bibinfo {year} {2018})}\BibitemShut {NoStop}%
\bibitem [{\citenamefont {Kolmogorov}(1963)}]{Kolmogorov1963Representation}%
  \BibitemOpen
  \bibfield  {author} {\bibinfo {author} {\bibfnamefont {A.~N.}\ \bibnamefont
  {Kolmogorov}},\ }\bibfield  {title} {\enquote {\bibinfo {title} {On the
  representation of continuous functions of many variables by superposition of
  continuous functions of one variable and addition},}\ }\href@noop {}
  {\bibfield  {journal} {\bibinfo  {journal} {Amer. Math. Soc. Transl}\
  }\textbf {\bibinfo {volume} {28}},\ \bibinfo {pages} {55} (\bibinfo {year}
  {1963})}\BibitemShut {NoStop}%
\bibitem [{\citenamefont {Le~Roux}\ and\ \citenamefont
  {Bengio}(2008)}]{Le2008Representational}%
  \BibitemOpen
  \bibfield  {author} {\bibinfo {author} {\bibfnamefont {N.}~\bibnamefont
  {Le~Roux}}\ and\ \bibinfo {author} {\bibfnamefont {Y.}~\bibnamefont
  {Bengio}},\ }\bibfield  {title} {\enquote {\bibinfo {title} {Representational
  power of restricted boltzmann machines and deep belief networks},}\ }\href
  {http://www.mitpressjournals.org/doi/abs/10.1162/neco.2008.04-07-510#.V8uLwrXG5Bw}
  {\bibfield  {journal} {\bibinfo  {journal} {Neural Comput.}\ }\textbf
  {\bibinfo {volume} {20}},\ \bibinfo {pages} {1631} (\bibinfo {year}
  {2008})}\BibitemShut {NoStop}%
\bibitem [{\citenamefont {Hornik}(1991)}]{Hornik1991Approximation}%
  \BibitemOpen
  \bibfield  {author} {\bibinfo {author} {\bibfnamefont {K.}~\bibnamefont
  {Hornik}},\ }\bibfield  {title} {\enquote {\bibinfo {title} {Approximation
  capabilities of multilayer feedforward networks},}\ }\href
  {http://www.sciencedirect.com/science/article/pii/089360809190009T}
  {\bibfield  {journal} {\bibinfo  {journal} {Neural networks}\ }\textbf
  {\bibinfo {volume} {4}},\ \bibinfo {pages} {251} (\bibinfo {year}
  {1991})}\BibitemShut {NoStop}%
\bibitem [{\citenamefont {Montakhab}\ and\ \citenamefont
  {Asadian}(2010)}]{PhysRevA.82.062313}%
  \BibitemOpen
  \bibfield  {author} {\bibinfo {author} {\bibfnamefont {A.}~\bibnamefont
  {Montakhab}}\ and\ \bibinfo {author} {\bibfnamefont {A.}~\bibnamefont
  {Asadian}},\ }\bibfield  {title} {\enquote {\bibinfo {title} {Multipartite
  entanglement and quantum phase transitions in the one-, two-, and
  three-dimensional transverse-field ising model},}\ }\href {\doibase
  10.1103/PhysRevA.82.062313} {\bibfield  {journal} {\bibinfo  {journal} {Phys.
  Rev. A}\ }\textbf {\bibinfo {volume} {82}},\ \bibinfo {pages} {062313}
  (\bibinfo {year} {2010})}\BibitemShut {NoStop}%
\bibitem [{\citenamefont {Chakrabarti}\ \emph {et~al.}(2008)\citenamefont
  {Chakrabarti}, \citenamefont {Dutta},\ and\ \citenamefont
  {Sen}}]{chakrabarti2008quantum}%
  \BibitemOpen
  \bibfield  {author} {\bibinfo {author} {\bibfnamefont {B.~K.}\ \bibnamefont
  {Chakrabarti}}, \bibinfo {author} {\bibfnamefont {A.}~\bibnamefont {Dutta}},
  \ and\ \bibinfo {author} {\bibfnamefont {P.}~\bibnamefont {Sen}},\
  }\href@noop {} {\emph {\bibinfo {title} {Quantum Ising phases and transitions
  in transverse Ising models}}},\ Vol.~\bibinfo {volume} {41}\ (\bibinfo
  {publisher} {Springer Science \& Business Media},\ \bibinfo {year}
  {2008})\BibitemShut {NoStop}%
\bibitem [{\citenamefont {Hinton}(2012)}]{Hinton2012}%
  \BibitemOpen
  \bibfield  {author} {\bibinfo {author} {\bibfnamefont {G.~E.}\ \bibnamefont
  {Hinton}},\ }\enquote {\bibinfo {title} {A practical guide to training
  restricted boltzmann machines},}\ in\ \href {\doibase
  10.1007/978-3-642-35289-8_32} {\emph {\bibinfo {booktitle} {Neural Networks:
  Tricks of the Trade: Second Edition}}},\ \bibinfo {editor} {edited by\
  \bibinfo {editor} {\bibfnamefont {G.}~\bibnamefont {Montavon}}, \bibinfo
  {editor} {\bibfnamefont {G.~B.}\ \bibnamefont {Orr}}, \ and\ \bibinfo
  {editor} {\bibfnamefont {K.-R.}\ \bibnamefont {M{\"u}ller}}}\ (\bibinfo
  {publisher} {Springer Berlin Heidelberg},\ \bibinfo {address} {Berlin,
  Heidelberg},\ \bibinfo {year} {2012})\ pp.\ \bibinfo {pages}
  {599--619}\BibitemShut {NoStop}%
\end{thebibliography}
%

\end{document}